\documentclass[11pt]{article}
\pdfoutput=1
\usepackage{jheppub}

\usepackage[...]{youngtab}
\usepackage{xfrac}
\usepackage{appendix}

\usepackage{amsfonts}
\usepackage{physics}
\usepackage[usenames,dvipsnames]{xcolor}
\usepackage{tabularx,multirow,array,amsmath,mathalfa}
\usepackage{color}
\usepackage{ccaption}
\usepackage{mathtools}
\usepackage{bbold}
\usepackage{kantlipsum}
\usepackage{slashbox}
\allowdisplaybreaks

\usepackage{changepage}
\usepackage[framemethod=TikZ]{mdframed}
\usepackage{lipsum}
\usepackage{titlesec}
\mdfdefinestyle{Frame-1}{%
    linecolor=black,
    outerlinewidth=2pt,
    roundcorner=20pt,
    innerrightmargin=15pt,
    innerleftmargin=15pt,
    backgroundcolor=gray!30!white}	
\newcommand\smallsection{%
  \titleformat*{\section}{\normalfont\small\bfseries}
	}
\newcommand{\frakg}{\ensuremath{\mathfrak{g}}}
\definecolor{Red}{rgb}{1,0,0}

\usepackage{graphicx}
\makeatletter
\newcommand\blfootnote[1]{%
  \begingroup
  \renewcommand\thefootnote{}\footnote{#1}%
  \addtocounter{footnote}{-1}%
  \endgroup
}
\newcommand*\bigcdot{\mathpalette\bigcdot@{.5}}
\newcommand*\bigcdot@[2]{\mathbin{\vcenter{\hbox{\scalebox{#2}{$\m@th#1\bullet$}}}}}
\makeatother
\usepackage{pgf}
\usepackage{tikz}
\usepackage[utf8]{inputenc}
\usetikzlibrary{arrows,automata}
\usetikzlibrary{positioning}
\tikzset{state/.style={rectangle, rounded corners, draw=black, very thick, minimum height=2em, inner sep=2pt, text centered,},}

\title{\boldmath Entanglement on linked boundaries in Chern-Simons theory with generic gauge groups}

\author[a,b]{Siddharth Dwivedi}
\author[b]{, Vivek Kumar Singh}
\author[b]{, Saswati Dhara}
\author[b]{, P. Ramadevi}
\author[c]{, Yang Zhou}
\author[b,d]{, Lata Kh Joshi}

\affiliation[a]{Center for Theoretical Physics, College of Physical Science and Technology, Sichuan University,\\
Chengdu, 610064, China}
\affiliation[b]{Department of Physics, Indian Institute of Technology Bombay, Mumbai 400076, India}
\affiliation[c]{Department of Physics and Center for Field Theory and Particle Physics, Fudan University, Shanghai 200433, China}
\affiliation[d]{Department of Theoretical Physics, 
Tata Institute of Fundamental Research, Mumbai 400005, India}

\abstract{We study the entanglement for a state on linked torus boundaries in $3d$ Chern-Simons theory with a generic gauge group and  present the asymptotic bounds of R\'enyi entropy at two different limits: (i) large Chern-Simons coupling $k$, and (ii)  large rank $r$ of the gauge group. These results show that the R\'enyi entropies cannot diverge faster than $\ln k$ and $\ln r$, respectively. We focus on torus links $T(2,2n)$ with topological linking number $n$. The R\'enyi entropy for these links shows a periodic structure in $n$ and vanishes whenever $n = 0 \text{ (mod } \textsf{p})$, where the integer \textsf{p} is a function of coupling $k$ and rank $r$. We highlight that the refined Chern-Simons link invariants can remove such a periodic structure in $n$.
}

\usepackage[normalem]{ulem}  

\begin{document} 
\begin{flushright} CTP-SCU/2017036, \\ TIFR/TH/17-41 \end{flushright}

\blfootnote{\textit{E-mail}: sdwivedi@scu.edu.cn, viveksingh@phy.iitb.ac.in, saswati123@phy.iitb.ac.in,  \\ ramadevi@phy.iitb.ac.in, yang\texttt{\_}zhou@fudan.edu.cn, latakj@theory.tifr.res.in }
	
		\vspace*{-2em}
	\maketitle
		\flushbottom

	\newpage


\section{Introduction}
\label{sec1}
Quantum entanglement is the characteristic nature of quantum systems. In the seminal work~\cite{Ryu:2006bv}, the holographic representation of the entanglement entropy measuring the amount of entanglement between a spatial region and its complement in conformal field theories was proposed in $AdS/CFT$ correspondence. This led to various works studying the entanglement structure from the holographic point of view \cite{Hubeny:2007xt, Lewkowycz:2013nqa, Bianchi:2012ev, Balasubramanian:2013rqa, Balasubramanian:2013lsa, Myers:2014jia, Swingle:2009bg, Nozaki:2013vta, Lashkari:2013koa, Faulkner:2013ica, Swingle:2014uza}. On the other hand, entanglement has also been studied in quantum field theories in various dimensions~\cite{Calabrese:2004eu, Calabrese:2009qy, Klebanov:2011uf, Casini:2009sr, Fursaev:2012mp}. It is generally an important question to understand the possible patterns of entanglement that can arise in field theory.

\begin{figure}[h]
\centerline{\includegraphics[width=4.4in]{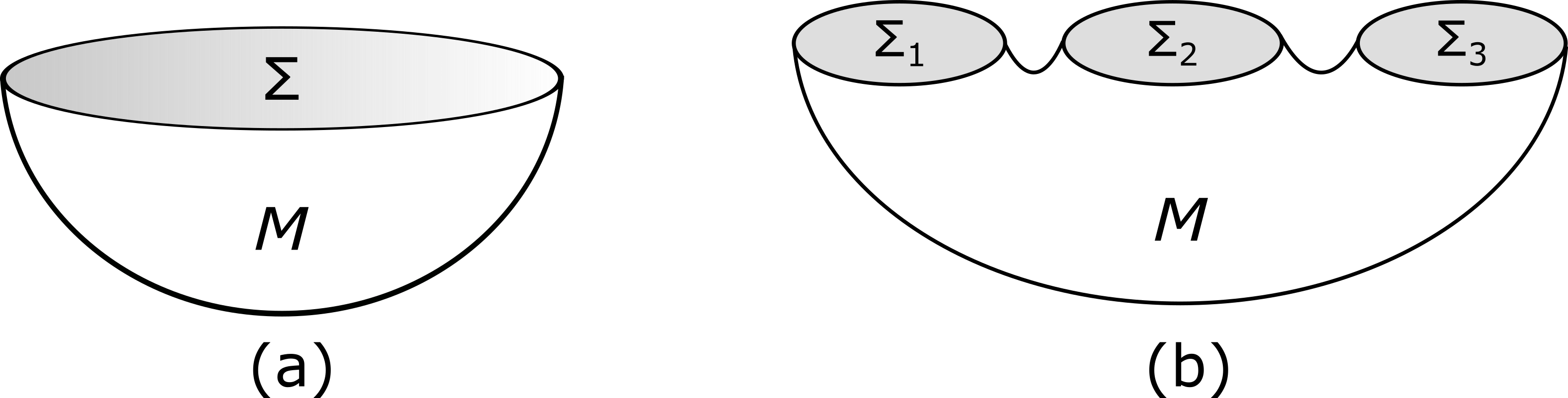}}
\caption[]{Figure (a) shows a manifold $M$ having a single boundary $\Sigma$. Figure (b) shows the manifold $M$ whose boundary has three components, i.e. $\Sigma_1 \cup \Sigma_2 \cup \Sigma_3$.}
\label{Multi-boundary}
\end{figure}

In particular, there has been a lot of interest to study entanglement in topological field theories for the simplicity. The best understood topological field theory is the 2+1 dimensional Chern-Simons theory \cite{Witten:1988hf}. The entanglement between connected spatial regions in Chern-Simons theories was obtained in \cite{Levin:2006zz, Dong:2008ft, Kitaev:2005dm}. In these works, there is a single boundary $\Sigma$ as shown in figure \ref{Multi-boundary}(a), which is then partitioned into two or more regions and one can study the entanglement between them. 

Another interesting problem is to consider the field theory on a manifold $M$ whose boundary $\Sigma$ consists of $n$ components $\Sigma = \Sigma_1 \cup \Sigma_2 \ldots \cup \Sigma_n$. For example, figure \ref{Multi-boundary}(b) shows a manifold having three boundary components. One can then study the entanglement structure between different components -- so called multi-boundary entanglement. 
In \cite{Balasubramanian:2014hda, Marolf:2015vma}, the multi-boundary entanglement was studied in the context of $AdS_3/CFT_2$, where the conformal field theory was defined on a Reimann surface having $n$ circle boundaries ($\Sigma = S^1 \cup S^1 \ldots \cup S^1$). The quantum state $\ket{\psi}$ was constructed by performing the Euclidean path integral on the Reimann surface and this state lived in the Hilbert space:
\begin{equation}
\ket{\psi} \in \mathcal{H}(S^1) \otimes \mathcal{H}(S^1) \otimes \ldots \otimes \mathcal{H}(S^1) ~,
\end{equation}
which is the tensor product of $n$ copies of $\mathcal{H}(S^1)$. Here $\mathcal{H}(S^1)$ denotes the Hilbert space associated to a single boundary circle. 
The holographic multi-boundary entanglement in this set-up can be obtained using the Ryu-Takayangi formula \cite{Ryu:2006bv}.

\begin{figure}[h]
\centerline{\includegraphics[width=2.4in]{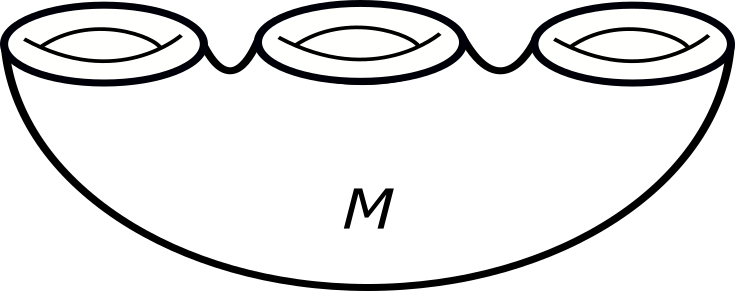}}
\caption[]{A manifold $M$ whose boundary $\Sigma$ has three components: $\Sigma=T^2 \cup T^2 \cup T^2$. Here the tori are unlinked, but it is straightforward  to construct boundaries where they are non-trivially linked. One can define a quantum state on this kind of boundaries and study its entanglement structure.}
\label{Multi-torus-boundary}
\end{figure}

Recently, the multi-boundary entanglement has been studied \cite{Salton:2016qpp, Balasubramanian:2016sro} for Chern-Simons theory on a three dimensional manifold $M$ whose boundary $\Sigma$ consists of $n$ topologically linked tori, i.e. $\Sigma = T^2 \cup T^2 \ldots \cup T^2$. For example, figure \ref{Multi-torus-boundary} shows a manifold whose boundary consists of three tori.
A quantum state $\ket{\psi}$ on such a boundary $\Sigma$ can be defined by performing the Chern-Simons path integral on manifold $M$. This state lives in the following Hilbert space:
\begin{equation}
\ket{\psi} \in \mathcal{H}(T^2) \otimes \mathcal{H}(T^2) \otimes \ldots \otimes \mathcal{H}(T^2) ~.
\end{equation}
Note that there can be many choices of the bulk manifold $M$ which satisfies the required boundary $\Sigma = T^2 \cup T^2 \ldots \cup T^2$. In \cite{Salton:2016qpp, Balasubramanian:2016sro}, a special manifold was selected which is called the link complement, which we will discuss briefly in section~\ref{sec2}. The state $\ket{\psi}$ is called the link state as it contains the topological information of the link. Thus, by a certain construction, the colored Jones invariant assigns a quantum entanglement structure to a link in a 3-sphere. This enables one to study knots and links in a information-theoretic approach. In \cite{Balasubramanian:2016sro}, the entanglement structure of this state was obtained for U(1) and SU(2) Chern-Simons theories at level $k$. Incidentally, the SU(2) entanglement entropy was shown to be more powerful than that of U(1) in  capturing the topological linking. Further, there are a class of links sharing same SU(2) but different SU($N$) invariants.  Hence, the theme of the present work is to explore the entanglement entropy and R\'enyi entropy for other compact gauge groups, which will provide richer information both for the links and for the non-abelian theories. 

In this note, we will follow \cite{Balasubramanian:2016sro} and focus on the R\'enyi entropy between two sub-links and its asymptotic bounds for large $k$ and/or large rank $r$ for various gauge groups. Specifically, we present the results for torus links $T(2,2n)$ with analytical results for $n=1$ and $n=2$. We do see interesting periodic structure of the R\'enyi entropy, whose periodicity is an integer \textsf{p} determined as a function of rank $r$
 and level $k$.

The paper is organized as follows. In section \ref{sec2}, we review the link state and the computation of reduced density matrix. In section \ref{sec3}, we study the upper bound of the R\'enyi entropy, which is given by the logarithm of the dimension of Hilbert space. We particularly discuss the asymptotic bounds at large $k$ limit and/or large rank limit. In section \ref{sec4}, we focus on the entanglement structure for a family of torus links having linking number $n$, which shows interesting periodic structure in $n$. We also show a potential refinement of the entanglement structures by working with a refined Chern-Simons invariants for links. We conclude and discuss future questions in section \ref{sec5}.

\section{Review of link invariants and reduced density matrix}
\label{sec2}
Witten, in the celebrated work \cite{Witten:1988hf}, argued that three-dimensional Chern-Simons theory on $S^3$ provides a natural framework to study the topological invariants of knots and links. In particular, the Jones polynomial for a link $\mathcal{L}$ can be realized as the Chern-Simons partition function of $S^3$ in the presence of Wilson line along $\mathcal{L}$, in the spin $\frac{1}{2}$ representation of $SU(2)$ gauge group. The explicit connection can be made if the complex variable $q$ in the Jones polynomial is taken as:
\begin{equation}
q = \text{exp}\left(\frac{2 \pi i}{k+2} \right) ~,
\end{equation}
where $k$ is the Chern-Simons coupling constant being a non-negative integer. One can generalize the invariants of links by computing the Chern-Simons partition functions for different gauge groups $G$ with Chern-Simons level $k$. For example, the $n$ colored Jones invariant is given by the Wilson loops in the spin $\frac{n}{2}$  representation of $SU(2)$ group, while the defining $N$-dimensional representations of $SU(N)$ and $SO(N)$, will give the HOMFLY-PT and Kauffman invariants, respectively. Generally, the complex variable $q = \text{exp}\left(\frac{2 \pi i}{k+y} \right)$, where $y$ is the dual Coxeter number of the gauge group.

\subsection{Link complement and link state}
\begin{figure}[h]
\centerline{\includegraphics[width=6.4in]{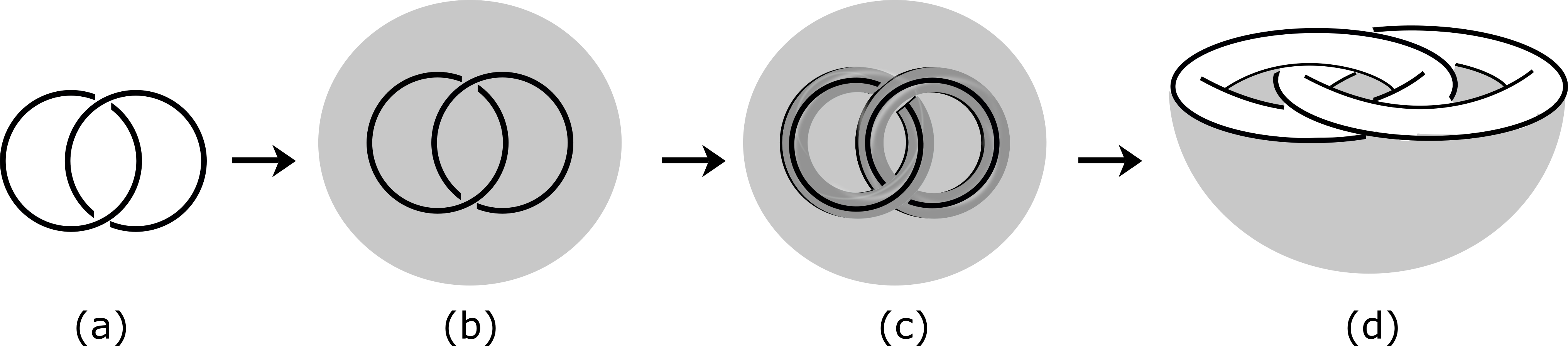}}
\caption[]{Figure (a) shows the Hopf link which is embedded into $S^3$ in figure (b). The Hopf link is first thickened into a tubular neighbourhood in figure (c) and then removed from $S^3$. The resulting manifold $S^3 \backslash \mathrm{H}$ is called the Hopf link complement as shown in figure (d) whose boundary consists of two linked tori $T^2 \cup T^2$. The Chern-Simons path integral gives a quantum state on this boundary whose entanglement structure can be studied.}
\label{Link-complement-Hopflink}
\end{figure}

Consider a link $\mathcal{L}$ in $S^3$, which has $p$ knot components. The link can have non-trivial linking between different knot components. Let us thicken the link $\mathcal{L}$ to obtain a tubular neighborhood, which is like $p$ linked solid tori centered about $\mathcal{L}$. Next we drill out this tubular neighborhood from $S^3$. By doing this, $S^3$ will split into two manifolds: one is the $p$ linked solid tori (say $\mathcal{L}_{\text{tub}}$) and the other is $S^3 \backslash \mathcal{L}_{\text{tub}}$,  which we simply denote as $S^3 \backslash \mathcal{L}$. We call the latter as the link complement. Note that both of these have the same boundary (but with opposite orientations), which is precisely $p$ linked torus. We illustrate this procedure explicitly in figure \ref{Link-complement-Hopflink} for an example of Hopf link ($\mathrm{H}$).

If we have a Wilson line in the bulk of a solid torus running parallel to its longitude cycle and carrying an integrable representation $a$, the Chern-Simons path integral will give a state $\ket{a}$ on the boundary. All possible integrable representations\footnote{In Appendix \ref{appendix-A}, we have briefly reviewed the integrable representations.} define this Hilbert space $\mathcal{H}_{T^2}$ on the boundary. For a given algebra $\frakg$ of rank $r$ and level $k$, the number $\mathcal{I}_{\frakg}$ of integrable representations is finite.  Thus we get a finite dimensional Hilbert space whose dimension is given by: 
\begin{equation}
\text{dim}(\mathcal{H}_{T^2}) = \mathcal{I}_{\frakg}(k) ~.
\label{dim=integrable}
\end{equation}
In the case of Hopf link complement $S^3 \backslash \mathrm{H}$, there are two torus boundaries as shown in figure \ref{Link-complement-Hopflink}(d). Thus the resulting Hilbert space, in this case, would be $\mathcal{H}_{T^2} \otimes \mathcal{H}_{T^2}$ and any state can be written as:
\begin{equation}
\ket{\psi} = \sum _{a} \sum _{b} C_{a b} \ket{a} \otimes \ket{b} \equiv \sum _{a,b}  C_{ab} \ket{a,b} ~,
\end{equation}
where $a$ and $b$ run over the possible integrable representations. The coefficients $C_{ab}$ are complex coefficients which depend upon the topological properties of $S^3 \backslash \mathrm{H}$ and are precisely the expectation value or the quantum invariants of the Hopf link embedded in $S^3$, with the two individual components of Hopf link carrying representations $a$ and $b$. 

This can be easily generalized to a link $\mathcal{L}$ made of $p$ knot components: $\mathcal{L} = \mathcal{K}_1 \cup \mathcal{K}_2 \cup \ldots \cup \mathcal{K}_p$. The link state, in this case, is given by:
\begin{equation}
\ket{\mathcal{L}} = \sum_{a_1,a_2,\ldots,a_p} C_{a_1, a_2, \ldots, a_p} \ket{a_1, a_2, \ldots, a_p} ~,
\label{link-state}
\end{equation}
where $a_i$ is the integrable representation associated to the $i\text{-th}$ knot (component) and $C_{\{a_i \}}$ is the link invariant of $\mathcal{L}$. The sum runs over all the integrable representations. 

\subsection{Reduced density matrix from link state}
For a general link $\mathcal{L}$ consisting of $p$ knot components, the total Hilbert space is given by,
\begin{equation}
\mathcal{H} = {\mathcal{H}_{T^2}}^{\otimes p} ~.
\label{n-copies-HS}
\end{equation}
One can bi-partition the total Hilbert space into two and trace out one of them. By this approach, one can measure the entanglement encoded in the global state $\ket{\mathcal{L}}$. We illustrate the procedure for a two-component link in the following. 

We start with the state given in eqn.(\ref{link-state}) for a two-component link $\mathcal{L}$ with link invariants $C_{ab}$. The total density matrix is given by:
\begin{equation}
\rho_{t} = \frac{\ket{\mathcal{L}} \bra{\mathcal{L}}}{\braket{\mathcal{L}}} ~,
\label{normalizing-rho}
\end{equation}
which obeys the pure state condition $\rho_{t}^2=\rho_{t}$. After tracing out one of the components, we get a reduced density matrix:
\begin{equation}
\rho  \equiv \sum_{b} \expval{\rho_{t}}{b} = \sum _{a, a^{\prime}} \left( \frac{\sum _{b} C_{ab} {C_{a^{\prime}b}}^{*} }{\sum _{c, d} C_{cd} {C_{cd}}^{*} } \right) \ket{a} \bra{a^{\prime}} ~,
\end{equation}
where $*$ represents the complex conjugate. Since the total Hilbert space is finite dimensional, the matrix $\rho$ is also finite dimensional with elements given as:
\begin{equation}
\rho  = \sum_{i,j=1}^{\mathcal{I}_{\frakg}} \rho_{ij} \ket{a_i} \bra{a_j} \quad;\quad \rho_{ij} = \frac{\sum_{l=1}^{\mathcal{I}_{\frakg}} C_{a_i, a_l} {C_{a_j, a_l}}^{*} }{\sum_{m=1}^{\mathcal{I}_{\frakg}} \sum_{n=1}^{\mathcal{I}_{\frakg}} C_{a_m, a_n} {C_{a_m, a_n}^{*}}} ~.
\label{rho-reduced}
\end{equation}
From here one can solve the eigenvalues of reduced density matrix and compute the R\'enyi entropy for any given link $\mathcal{L}$.
\section{R\'enyi entropy for link state associated with a link $\mathcal{L}$} \label{sec3}
In this section, we will study some general features of the R\'enyi entropy of link state $\ket{\mathcal{L}}$, where $\mathcal{L}$ is made of $p$ knot components. We take the Chern-Simons theory with group $G$ of rank $r$ and the Chern-Simons level as $k$. The total Hilbert space is product of $p$ copies of $\mathcal{H}_{T^2}$ as in eqn.(\ref{n-copies-HS}). Consider the bi-partition $(m | p-m)$ of this Hilbert space ($m \leq p-m$):
\begin{equation}
\mathcal{H} = ({\mathcal{H}_{T^2}}^{\otimes m}) \otimes ({\mathcal{H}_{T^2}}^{\otimes p-m})
\end{equation}  
The entanglement structure of link state $\ket{\mathcal{L}}$ can be studied by tracing out ${\mathcal{H}_{T^2}}^{\otimes p-m}$ and the resulting reduced density matrix is of order $\mathcal{I}_{\frakg}^m$, where $\mathcal{I}_{\frakg}$ is the number of integrable representations\footnote{In Appendix \ref{appendix-A}, we have given the number of integrable representations for various classical and exceptional Lie algebras.} of the affine algebra $\frakg_k$. The R\'enyi entropy therefore is given as ($\alpha \geq 0$):
\begin{equation}
\mathcal{R}_{\alpha}(\mathcal{L}) = \frac{1}{1-\alpha} \ln \text{tr}(\rho^{\alpha}) = \frac{1}{1-\alpha} \ln (\lambda_1^{\alpha} + \lambda_2^{\alpha} + \ldots + \lambda_{\mathcal{I}_{\frakg}^m}^{\alpha}) ~,
\end{equation}
where $\lambda_i$ are the eigenvalues of reduced density matrix. When $\alpha \to 0$, the R\'enyi entropy is maximum, which is $\ln \mathcal{I}_{\frakg}^m$ and when $\alpha \to \infty$, the R\'enyi entropy reaches a minimum value, which is $\mathcal{R}_{\infty} = \ln \left(\lambda_{\textsf{max}}\right)^{-1}$, where $\lambda_{\textsf{max}}$ is the maximum of all the eigenvalues of density matrix:
\begin{equation}
\lambda_{\textsf{max}} = \text{max} \left\{ \lambda_{1}, \lambda_{2}, \ldots, \lambda_{\mathcal{I}_{\frakg}^m} \right\}  ~.
\end{equation}
Thus the R\'enyi entropy $\mathcal{R}_{\alpha}$ is bounded:
\begin{equation}
0 \leq \mathcal{R}_{\infty}(\mathcal{L})  \leq  \mathcal{R}_{\alpha}(\mathcal{L}) \leq m\ln \mathcal{I}_{\frakg}  ~.
\label{RE-bound}
\end{equation}
In the following, we attempt the asymptotic bounds of $\mathcal{R}_{\alpha}$ when: (i) the level $k$ is large, (ii) the rank $r$ is large and (iii) both $k$ and $r$ are large but their ratio  is finite. From now on, we will set $m=1$, but the generalization to higher integer $m$ is straightforward.

\subsection{Upper bounds when the level $k \to \infty$}
Now we study the asymptotic behavior of $\ln \mathcal{I}_{\frakg}$ as $k \to \infty$. The generating function for counting the number of integrable representations are given in  Appendix \ref{appendix-A} (see eqn.(\ref{Integrable-rep-arbit-algebra})). For the well-known Lie algebras, we find that $\ln \mathcal{I}_{\frakg} \sim \ln k$, for large values of $k$. One can obtain the following expansion for various classical and exceptional Lie algebras of finite rank, by expanding $\ln \mathcal{I}_{\frakg}$ as series in $k^{-1}$:
\begin{align}
\ln \mathcal{I}_{\mathfrak{su}(N)} &= \ln\left( \frac{k^{N-1}}{(N-1)!} \right) + \frac{N(N-1)}{2k} + \mathcal{O}\left( \frac{1}{k^2} \right) + \ldots \nonumber \\[10pt]
\ln \mathcal{I}_{\mathfrak{so}(2N+1)} &= \begin{cases}
    \ln k + \dfrac{1}{k} + \mathcal{O}\left( \dfrac{1}{k^2} \right) + \ldots & \text{for $\mathfrak{so}(3)$} \\[10pt]
		N\ln k + \ln\left( \dfrac{2^{2-N}}{N!} \right) + \mathcal{O}\left( \dfrac{1}{k} \right)  + \ldots & \text{for $N \geq 2$}
\end{cases}
 \nonumber \\[10pt]
\ln \mathcal{I}_{\mathfrak{sp}(2N)}  &= \ln\left( \frac{k^{N}}{N\,!} \right) + \frac{N(N+1)}{2k} + \mathcal{O}\left( \frac{1}{k^2} \right) + \ldots \nonumber \\[10pt]
\ln \mathcal{I}_{\mathfrak{so}(2N)} &= N\ln k + \ln\left( \dfrac{2^{3-N}}{N!} \right) + \mathcal{O}\left( \dfrac{1}{k} \right) + \ldots \nonumber \\[10pt]
\ln \mathcal{I}_{E_6}  &= \ln k^{6} -\ln 17280 + \mathcal{O}\left( \frac{1}{k} \right) + \ldots \nonumber \\
\ln \mathcal{I}_{E_7}  &= \ln k^{7} -\ln 1451520 + \mathcal{O}\left( \frac{1}{k} \right) + \ldots \nonumber \\
\ln \mathcal{I}_{E_8}  &= \ln k^{8} -\ln 696729600 + \mathcal{O}\left( \frac{1}{k} \right) + \ldots \nonumber \\
\ln \mathcal{I}_{F_4}  &= \ln k^{4} -\ln 288 + \mathcal{O}\left( \frac{1}{k} \right) + \ldots \nonumber \\
\ln \mathcal{I}_{G_2}  &= \ln k^{2} -\ln4 + \mathcal{O}\left( \frac{1}{k} \right) + \ldots ~.
\label{Integrable-rep-k-expansion}
\end{align}
From the above expansions, we observe that $\ln \mathcal{I}_{\frakg}$ diverges as $\ln k$ at the large $k$ limit with a fixed coefficient:
\begin{align}
\Aboxed{\lim_{k \to \infty} \left( \frac{\ln \mathcal{I}_{\frakg} }{\ln k} \right) = r} ~.
\end{align}
This result suggests that we must study the asymptotic bounds of the ratio of R\'enyi entropy and $\ln k$. For this, we divide eqn.(\ref{RE-bound}) by $\ln k$ and take the limit $k \to \infty$. So, we will get (for $m=1$ case):
\begin{align}
\Aboxed{\lim_{k \to \infty} \left( \frac{\mathcal{R}_{\alpha}(\mathcal{L})}{\ln k} \right) = c_1(\alpha) \quad;\quad 0 \leq \lim_{k \to \infty} \left( \frac{\mathcal{R}_{\infty}(\mathcal{L})}{\ln k} \right) \leq c_1(\alpha) \leq r} ~,
\label{bound-RE-limit-k-infinity}
\end{align}
where $c_1(\alpha) \geq 0$ is a constant for a given algebra of finite rank. We will show explicit results of $\mathcal{R}_{\alpha}(\mathcal{L})$ for the simplest link Hopf link, as well as the torus link $T(2,4)$, later in this section, which is consistent with this observation.

\subsection{Upper bounds when the rank $r \to \infty$}
Motivated by the rank-level duality in Chern-Simons theory, we may consider the asymptotic expansion of $\ln \mathcal{I}_{\frakg}$ when the rank $r$ is large but $k$ is fixed. Indeed, we get $\ln \mathcal{I}_{\frakg} \sim \ln r$ as expected and up to first order in $r^{-1}$, the expansion of $\ln \mathcal{I}_{\frakg}$ can be written for various classical Lie groups as follows:
\begin{align}
\ln \mathcal{I}_{\mathfrak{su}(r)}  &= k\ln r - \ln(k!) + \dfrac{k(k-1)}{2r} + \mathcal{O}\left( \frac{1}{r^2} \right) + \ldots \nonumber \\[10pt]
\ln \mathcal{I}_{\mathfrak{so}(2r+1)} &= \begin{cases}
    \dfrac{k}{2}\ln r - \ln(\frac{k}{2}!) + \dfrac{k (k+14)}{8r}   + \ldots, & \text{for even $k$} \\[10pt]
    \left(\dfrac{k-1}{2}\right)\ln r + \ln 3 - \ln(\frac{k-1}{2}!) + \dfrac{(k-1) (3k+7)}{24r}  + \ldots, & \text{for odd $k$}
\end{cases}
 \nonumber \\[10pt]
\ln \mathcal{I}_{\mathfrak{sp}(2r)}  &= k\ln r - \ln(k!) + \dfrac{k(k+1)}{2r} + \mathcal{O}\left( \dfrac{1}{r^2} \right) + \ldots \nonumber \\[10pt]
\ln \mathcal{I}_{\mathfrak{so}(2r)}  &= \begin{cases}
    \dfrac{k}{2}\ln r - \ln(\frac{k}{2}!) + \dfrac{k (k+26)}{8r}   + \ldots, & \text{for even $k$} \\[10pt]
    \left(\dfrac{k-1}{2}\right)\ln r + \ln 4 - \ln(\frac{k-1}{2}!) + \dfrac{(k-1) (k+5)}{8r}  + \ldots, & \text{for odd $k$}~.
\end{cases} \nonumber
\end{align}
We can see that $\ln \mathcal{I}_{\frakg}$ diverges as $\ln r$ as $r \to \infty$ for a finite level and we can write the following limit for a given value of level $k$:
\begin{align}
\Aboxed{\lim_{r \to \infty} \left( \frac{\ln \mathcal{I}_{\frakg} }{\ln r} \right) = f_{\frakg}(k)} ~.
\end{align}
Here the function $f_{\frakg}(k)$ depends on the choice of algebra and for classical Lie algebras, we deduce:
\begin{equation}
f_{\frakg}(k) = \begin{cases}
  k, & \text{for $\frakg = \mathfrak{su}(\infty),\, \mathfrak{sp}(\infty)$} \\[5pt]
	\left\lfloor{\dfrac{k}{2}}\right\rfloor \quad & \text{, \quad for $\frakg = \mathfrak{so}(\infty)$} ~,
 \end{cases} 
\end{equation}
where $\left\lfloor{x}\right\rfloor$ denotes the largest integer less than or equal to $x$. Thus the R\'enyi entropy in the limit of rank $r \to \infty$, for a given value of $k$, can be written as:
\begin{align}
\Aboxed{\lim_{r \to \infty} \left( \frac{\mathcal{R}_{\alpha}(\mathcal{L})}{\ln r} \right) = c_2(\alpha) \quad;\quad 0 \leq \lim_{r \to \infty} \left( \frac{\mathcal{R}_{\infty}(\mathcal{L})}{\ln r} \right) \leq c_2(\alpha) \leq f_{\frakg}(k)} ~,
\label{bound-RE-limit-r-infinity}
\end{align}
where $c_2(\alpha) \geq 0$ is a constant for a given group and a given level. Now it will be interesting to see the asymptotic behavior of upper bound, when both the rank and the level are taken to be large. 

\subsection{Upper bounds in the double scaling limit}
In the double scaling limit, we take both rank $r \to \infty$ and level $k \to \infty$ such that the generalized ratio is kept finite: 
\begin{equation}
\lim_{\substack{r \to \infty \\ k \to \infty}} \left( \dfrac{r^a}{k^b} \right) = c ~,
\label{general-ratio}
\end{equation}
where $a,b,c \in \mathbb{R}_{+}$ are constants. In such a case, $\ln \mathcal{I}_{\frakg}$ shows different asymptotic behavior depending upon the ratio of $a$ and $b$ which is clear from the following limits for classical Lie algebras:
\begin{align}
a < b &: \begin{cases}
    \lim_{\substack{r \to \infty \\ k \to \infty}} \left( \dfrac{\ln \mathcal{I}_{\frakg}}{k \ln k} \right) = \dfrac{b-a}{a}  \quad, & \text{for $\frakg = \mathfrak{su}(\infty), \, \mathfrak{sp}(\infty)$} \\[10pt]
   \lim_{\substack{r \to \infty \\ k \to \infty}} \left( \dfrac{\ln \mathcal{I}_{\frakg}}{k \ln k} \right) = \dfrac{b-a}{2a}  \quad, & \text{for $\frakg = \mathfrak{so}(\infty)$}
\end{cases} \nonumber \\[10pt]
a = b &: \begin{cases}
    \lim_{\substack{r \to \infty \\ k \to \infty}} \left( \dfrac{\ln \mathcal{I}_{\frakg}}{k} \right) = \text{\large{ln}} \left( \dfrac{(1+c^{1/a})^{1+c^{1/a}}}{(c^{1/a})^{c^{1/a}}}\right)  \quad, & \text{for $\frakg = \mathfrak{su}(\infty), \, \mathfrak{sp}(\infty)$} \\[10pt]
   \lim_{\substack{r \to \infty \\ k \to \infty}} \left( \dfrac{\ln \mathcal{I}_{\frakg}}{k} \right) = \text{\large{ln}}\left( \dfrac{(1+2c^{1/a})^{\frac{1}{2}+c^{1/a}}}{(2c^{1/a})^{c^{1/a}}}\right)  \quad, & \text{for $\frakg = \mathfrak{so}(\infty)$}
\end{cases} \nonumber \\[10pt]
a > b &: 
    \lim_{\substack{r \to \infty \\ k \to \infty}} \left( \dfrac{\ln \mathcal{I}_{\frakg}}{k^b \ln k} \right) = \dfrac{(a-b)c^{1/a}}{a} \quad,\quad \text{for $\frakg = \mathfrak{su}(\infty), \, \mathfrak{sp}(\infty), \, \mathfrak{so}(\infty)$} ~.
		\label{Integrable-rep-k+r-infinity}
\end{align}
The various asymptotic bounds of R\'enyi entropy therefore can be given as: 
\begin{align}
a < b &: \quad 0 \leq \lim_{\substack{r \to \infty \\ k \to \infty}} \left(\frac{\mathcal{R}_{\infty}}{k \ln k}\right) \leq \lim_{\substack{r \to \infty \\ k \to \infty}} \left(\frac{\mathcal{R}_{\alpha}}{k \ln k}\right) \leq  \lim_{\substack{r \to \infty \\ k \to \infty}} \left( \dfrac{\ln \mathcal{I}_{\frakg}}{k \ln k} \right) \nonumber \\[5pt]
a = b &: \quad 0 \leq \lim_{\substack{r \to \infty \\ k \to \infty}} \left(\frac{\mathcal{R}_{\infty}}{k}\right) \leq \lim_{\substack{r \to \infty \\ k \to \infty}} \left(\frac{\mathcal{R}_{\alpha}}{k}\right) \leq  \lim_{\substack{r \to \infty \\ k \to \infty}} \left( \dfrac{\ln \mathcal{I}_{\frakg}}{k} \right) \nonumber \\[5pt]
a > b &: \quad 0 \leq \lim_{\substack{r \to \infty \\ k \to \infty}} \left(\frac{\mathcal{R}_{\infty}}{k^b \ln k}\right) \leq \lim_{\substack{r \to \infty \\ k \to \infty}} \left(\frac{\mathcal{R}_{\alpha}}{k^b \ln k}\right) \leq  \lim_{\substack{r \to \infty \\ k \to \infty}} \left( \dfrac{\ln \mathcal{I}_{\frakg}}{k^b \ln k} \right) ~.
	\label{RE-limit-k+r-infinity}
\end{align}
The upper bounds for all these cases are given in eqn.(\ref{Integrable-rep-k+r-infinity}). In the following, we illustrate the various asymptotics by giving two examples of Hopf link ($\mathrm{H}$) and $T(2,4)$ link, which are two-component links.

\subsection{Example-1: R\'enyi entropy for Hopf link state for generic group}
For the simplest non-trivial link, i.e. Hopf link, the link state is given as:
\begin{equation}
\ket{\mathrm{H}} = \sum _{a} \sum _{b} \mathcal{S}_{ab} \left|a,b\right\rangle ~,
\label{Hopflink-state}
\end{equation}
where, $\mathcal{S}_{ab}$ is the Hopf link invariant with $\mathcal{S}$ (eqn. (\ref{S-T-Weyl-definition})) being a generator of modular group SL($2, \mathbb{Z}$) of torus diffeomorphism. With this link state, the reduced density matrix elements corresponding to a bi-partition $(1|1)$ can be computed (see for example \cite{Balasubramanian:2016sro}):
\begin{equation}
\rho_{ij} = \frac{1}{\mathcal{I}_{\frakg}} \delta_{ij} ~,
\end{equation}
Thus the R\'enyi entropy is maximum and is equal to the upper bound of eqn.(\ref{RE-bound}) with $m=1$:
\begin{equation}
\mathcal{R}_{\alpha}^{G}(\mathrm{H}) = \ln \mathcal{I}_{\frakg} ~.
\end{equation}
Observing the large $k$ expansion of $\ln \mathcal{I}_{\frakg}$ for various algebras in eqn.(\ref{Integrable-rep-k-expansion}), the R\'enyi entropy for Hopf link can be written as following:
\begin{equation}
\boxed{\mathcal{R}_{\alpha}^{G}(\mathrm{H}) = \ln \mathcal{I}_{\frakg} = r \ln k - \ln r! - \ln x(\frakg)  +\mathcal{O}\left(\frac{1}{k} \right) + \ldots} ~,
\label{Integrable-expansion-g}
\end{equation}
where $r$ is the finite rank of the algebra $\frakg$. The value of constant term $\ln {x(\frakg)}$, depends explicitly on the type of algebra. This term can be given an interpretation of the asymptotic deviation of $\ln \mathcal{I}_{\frakg}$ computed for an algebra $\frakg$, from $\ln \mathcal{I}_{\frakg_0}$, which is computed for the algebra $\frakg_0$ having a maximum number of integrable representations, with both the algebras having the same rank. From eqn.(\ref{Integrable-rep-arbit-algebra}), it is clear that for a given rank and level, $\mathcal{I}_{\frakg}$ will be maximum if all the dual Kac labels are unity (the classical Lie algebras $A_N$ and $C_N$ has this property). The number of integrable representations in such cases is given as\footnote{The $\mathfrak{so}(4)$ algebra with rank 2 is an exception for which the number of integrable representations is greater than that given in eqn.(\ref{Integrable-rep-g0}). Its integrability condition is given in eqn.(\ref{Number-Integrability-BN}).}:
\begin{equation}
\mathcal{I}_{\frakg_0} = \dfrac{(k+r)!}{r!\,k!} ~,
\label{Integrable-rep-g0}
\end{equation}
and we can write the asymptotic expansion at large level as:
\begin{equation}
\mathcal{R}_{\alpha}^{G_0}(\mathrm{H}) = \ln \mathcal{I}_{\frakg_0} = r \ln k -\ln r!  +\mathcal{O}\left(\frac{1}{k} \right) + \ldots ~,
\label{Integrable-expansion-g0}
\end{equation}
where $G_0$ is the group corresponding to $\frakg_0$. Now if we take the difference between eqn.(\ref{Integrable-expansion-g}) and eqn.(\ref{Integrable-expansion-g0}) and take the limit $k \to \infty$, we will get:
\begin{equation}
\lim_{k \to \infty}(\ln \mathcal{I}_{\frakg} - \ln \mathcal{I}_{\frakg_0}) \equiv \lim_{k \to \infty}\Delta \ln \mathcal{I}_{\frakg} = \lim_{k \to \infty} \Delta \mathcal{R}_{\alpha}^{G}(\mathrm{H})  = \ln x(\frakg)^{-1} ~.
\end{equation}
It is interesting to note that the quantity $\ln x(\frakg)$ converges for all the well known Lie algebras:
\begin{align}
\ln x(\frakg) = 
\begin{cases}
    0  & \text{,\,\, for $\frakg=\mathfrak{su}(r), \,\mathfrak{sp}(r)$} \\[0pt]
		0  & \text{,\,\, for $\frakg=\mathfrak{so}(3), \,\mathfrak{so}(5), \,\mathfrak{so}(6)$} \\[0pt]
    \ln \left( 2^{\left\lceil{\sfrac{r}{2}}\right\rceil -3} \right) \quad & \text{,\,\, for $\frakg=\mathfrak{so}(r), \,r=4 \text{ and } r \geq 7$} \\[0pt]
		\ln 2^7 + \ln3^3 + \ln5  & \text{,\,\, for $\frakg=E_8$} \\[0pt]
		\ln 2^5 + \ln3^2  & \text{,\,\, for $\frakg=E_7$} \\[0pt]
		\ln 2^3 + \ln3  & \text{,\,\, for $\frakg=E_6$} \\[0pt]
		\ln 2^2 + \ln3  & \text{,\,\, for $\frakg=F_4$} \\[0pt]
		\ln 2  & \text{,\,\, for $\frakg=G_2$} ~,
\end{cases}
\end{align}
where $\left\lceil{\frac{r}{2}}\right\rceil$ denotes the smallest integer greater than or equal to $\frac{r}{2}$. 

\subsection{Example-2: R\'enyi entropy for $T(2,4)$ link state for SU(2) group}
The second example we discuss is the $T(2,4)$ link having linking number two and is made of two circles linked twice as shown in figure \ref{T(2,4)}.
\begin{figure}[h]
\centerline{\includegraphics[width=1.4in]{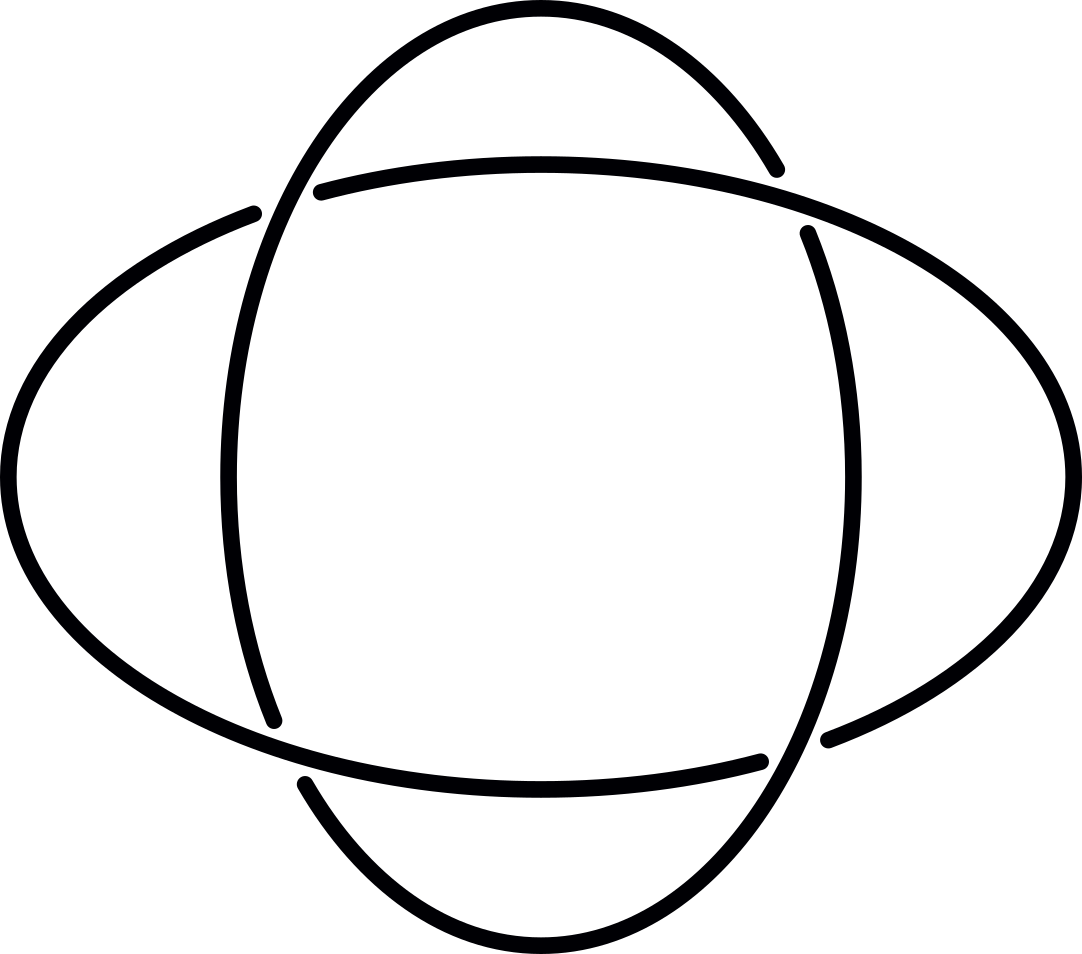}}
\caption[]{The $T(2,4)$ link.}
\label{T(2,4)}
\end{figure}
Actually, this is a member of a more general family of links $T(2,2n)$, which we will discuss in section \ref{sec4}. The link invariant and reduced density matrix of $T(2,4)$ can be obtained from eqn.(\ref{T(2,2n)-invariant}) and eqn.(\ref{T(2,2n)-density-matrix}) respectively, by substituting $n=2$. Using SU(2) invariant for this link, we obtained a closed form expression for eigenvalues of the reduced density matrix. The non-zero eigenvalues are:
\begin{equation}
T(2,4): 
\begin{cases}
   \lambda_{m}^{\text{SU}(2)} = \dfrac{2 \csc ^2\left(\dfrac{\pi -2 \pi  m}{2 k+4}\right)}{(k+2)^2} ; \, m=1,\ldots,\frac{k}{2}+1, & \text{for even $k$} \\[14pt]
    \lambda_{m}^{\text{SU}(2)} = \dfrac{2 \csc ^2\left(\dfrac{\pi -2 \pi  m}{2 k+4}\right)}{(k+1) (k+3)} ; \, m=1,\ldots,\frac{k+1}{2},  & \text{for odd $k$}
\end{cases} ~.
\label{T(2,4)-eigen-values}
\end{equation}
From these eigenvalues, it is difficult to write a closed form expression of the R\'enyi entropy for a finite value of $k$. However, in the limiting case of $k \to \infty$, one can obtain the order $\alpha$ R\'enyi entropy as following:
\begin{equation}
\boxed{
T(2,4): \quad \lim_{k \to \infty} \mathcal{R}_{\alpha}^{\text{SU}(2)} = \dfrac{\ln \left(4^{\alpha}-1\right) + \alpha  \ln \left(2/\pi^2\right) + \ln \zeta(2\alpha)}{1-\alpha}} ~,
\label{Renyi-T(2,4)-large-k}
\end{equation}
where $\zeta(2\alpha)$ is the Riemann zeta function evaluated at $2\alpha$. When $\alpha > 0$, the R\'enyi entropy converges as $k \to \infty$ which is evident from eqn.(\ref{Renyi-T(2,4)-large-k}). For $\alpha \to 1$ and $\alpha \to \infty$, the values of entanglement entropy and minimum R\'enyi entropy in the limit of $k \to \infty$ can be obtained as:
\begin{equation}
\boxed{T(2,4): 
\begin{cases}
   \lim_{k \to \infty} \mathcal{R}_{1}^{\text{SU}(2)} = 24\ln A - 2 \gamma -(17/3) \ln 2 \approx 0.887842 \\[5pt]
   \lim_{k \to \infty} \mathcal{R}_{\infty}^{\text{SU}(2)} = \ln \left( \dfrac{\pi^2}{8} \right) \approx 0.210018
\end{cases}} ~,
\end{equation}
where $A$ and $\gamma$ are the Glaisher's constant and Euler's constant respectively. 

We have seen that the R\'enyi entropy for Hopf link is maximum. Hence, for completeness, it would be interesting to compare R\'enyi entropy of the link $T(2,4)$ with respect to that of Hopf link, which is encoded in a quantity called R\'enyi divergence. We discuss the details in the following subsection.

\subsection{R\'enyi divergence $\mathcal{D}_{\alpha}(\mathcal{L} \parallel \mathrm{H})$}
The R\'enyi divergence $\mathcal{D}_{\alpha}$ of order $\alpha$ ($\alpha > 0$ and $\alpha \neq 1$), of the eigenvalue distribution $\{ \lambda_i \}$  of reduced density matrix for a two-component link state $\ket{\mathcal{L}}$, from the eigenvalue distribution $\{ \tilde{\lambda}_i \}$  of that of Hopf link state $\ket{\mathrm{H}}$, is defined as:
\begin{equation}
\mathcal{D}_{\alpha}(\mathcal{L} \parallel \mathrm{H}) = \frac{1}{\alpha-1} \ln \left( \sum_{i=1}^{\mathcal{I}_{\frakg}} \frac{\lambda_i^{\alpha}}{{\tilde{\lambda}}_i^{\alpha-1}} \right) ~.
\end{equation}
Now the R\'enyi entropy for Hopf link is maximum, implying that the eigenvalues of the density matrix are all equal:
\begin{equation}
\lambda_i = \frac{1}{\mathcal{I}_{\frakg}} \quad;\quad i=1,2,\ldots, \mathcal{I}_{g} ~.
\end{equation}
Using this, we get:
\begin{equation}
\mathcal{D}_{\alpha}(\mathcal{L} \parallel \mathrm{H}) = \ln \mathcal{I}_{\frakg} - \mathcal{R}_{\alpha}(\mathcal{L}) ~,
\end{equation}
where $\mathcal{R}_{\alpha}(\mathcal{L})$ is the R\'enyi entropy for link state $\ket{\mathcal{L}}$. Using the bound in eqn.(\ref{RE-bound}) for the R\'enyi entropy of a general two-component link $\mathcal{L}$, we can write the bound for this R\'enyi divergence as:
\begin{equation}
0 \leq \mathcal{D}_{\alpha}(\mathcal{L} \parallel \mathrm{H}) \leq \ln \mathcal{I}_{\frakg} - \mathcal{R}_{\infty}(\mathcal{L}) ~.
\end{equation}
Thus the upper and lower bounds of R\'enyi divergence $\mathcal{D}_{\alpha}$ are simply lowered by an amount $\mathcal{R}_{\infty}$ as compared to that of the bounds of R\'enyi entropy $\mathcal{R}_{\alpha}(\mathcal{L})$ in eqn.(\ref{RE-bound}). One can verify that for $\mathcal{L} = \mathrm{H}$, we get $\mathcal{D}_{\alpha} = 0$. For $\mathcal{L} = T(2,4)$ and group SU(2), the following bound for R\'enyi divergence can be given:
\begin{equation}
0 \leq \mathcal{D}_{\alpha}(T(2,4) \parallel \mathrm{H}) \leq \begin{cases}
   \ln(\dfrac{2k+2}{k^2+4k+4}) - 2\ln \left(\sin(\frac{\pi}{2k+4}) \right) \quad & \text{, \quad for even $k$} \\[10pt]
   \ln(\dfrac{2}{k+3}) - 2\ln \left(\sin(\frac{\pi}{2k+4}) \right) \quad & \text{, \quad for odd $k$}
\end{cases} ~.
\end{equation}
Ideally, we should get the analytic results for $R_{\alpha}(\mathcal{L})$ and $\mathcal{D}_{\alpha}(\mathcal{L} \parallel \mathrm{H})$ and their asymptotic bounds for two-component links. But unfortunately, finding the closed form expressions for the eigenvalues of density matrix is difficult. So in the following section, we attempt this issue numerically for the class of torus links $T(2,2n)$.
\section{Entanglement structure of link state on boundary of $S^3 \backslash T(2,2n)$}
\label{sec4}
In this section, we will study the entanglement structure of the link state defined on the two torus boundaries of $T(2,2n)$ link complement. This family of links consists of two unknots which are twisted and linked $n$ times, as shown in figure \ref{T(2,2n)}. 
\begin{figure}[h]
\centerline{\includegraphics[width=2.5in]{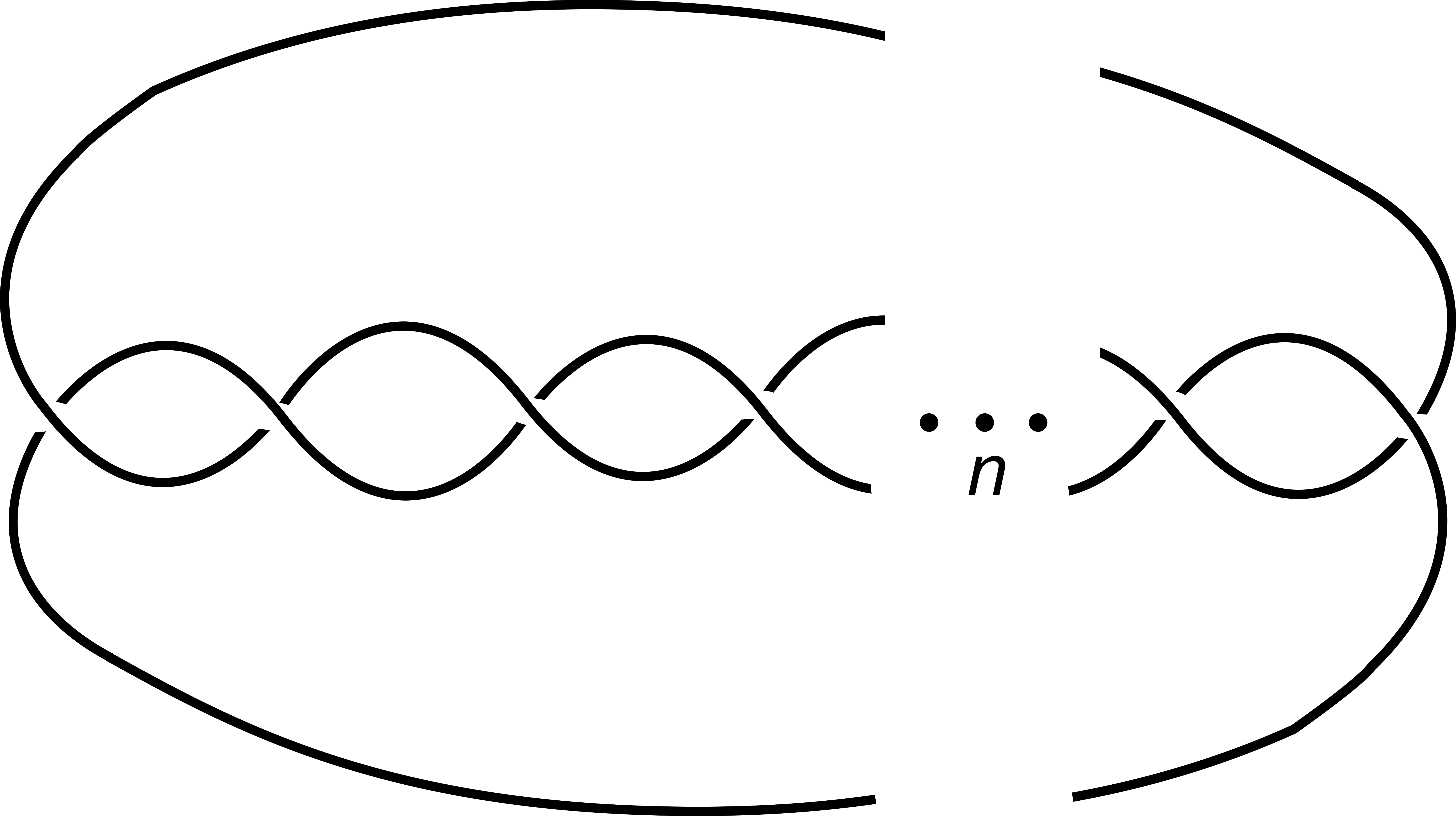}}
\caption[]{The $T(2,2n)$ link, where the two circles are twisted $n$ number of times. There are $2n$ crossings and the linking number is $n$.}
\label{T(2,2n)}
\end{figure}
The Hopf link is a member of this family with $n=1$. Using the results of \cite{Balasubramanian:2016sro}, the R\'enyi entropy of $T(2,2n)$ link state can be given for U(1) group as following:
\begin{equation}
\mathcal{R}_{\alpha} \left( n \right) = \ln \left(\frac{k}{\text{gcd}(k,n)} \right)  ~.
\end{equation}
Clearly, the entanglement structure is periodic in $n$ with period $k$ and the R\'enyi entropy vanishes when $n$ is a multiple of $k$:
\begin{equation}
\mathcal{R}_{\alpha} \left( n + k\mathbb{Z} \right) = \mathcal{R}_{\alpha} \left( n \right) \quad;\quad \mathcal{R}_{\alpha} \left( k \right) =0  ~.
\end{equation}
In fact, as we show in this section, this is a generic feature of $T(2,2n)$ link state and for any given group, the periodicity depends on $k$ as well as rank $r$ of the group: 
\begin{equation}
\mathcal{R}_{\alpha} \left( n + \textsf{p}(k,r)\mathbb{Z} \right) = \mathcal{R}_{\alpha} \left( n \right) \quad;\quad \mathcal{R}_{\alpha} \left( \textsf{p} \right) =0  ~,
\end{equation}
where $\textsf{p}$ is the fundamental period depending on the rank and the level. In the following, we discuss this structure in more detail.

The topological invariant for these links can be obtained using the Dehn surgery discussed in \cite{Witten:1988hf} and can be given in terms of $\mathcal{S}$ and $\mathcal{T}$ generators of modular group SL($2, \mathbb{Z}$) of torus diffeomorphism \cite{Balasubramanian:2016sro}:
\begin{equation}
C_{ab} =  \sum_{m} \frac{\left(\mathcal{S}\mathcal{T}^n \mathcal{S}\right)_{0m}}{\mathcal{S}_{0m}}\mathcal{S}_{am}\mathcal{S}_{bm} ~.
\label{T(2,2n)-invariant}
\end{equation}
Here $\mathcal{T}^n$ creates the required $n$-fold Dehn-twist in the torus resulting in the $n$ times twisting of the two circles. The $\mathcal{T}$-matrix elements are given as:
\begin{equation}
\mathcal{T}_{ab} = \text{exp}\left( 2 \pi i \, \frac{\Lambda(a) \bigcdot (\Lambda(a) + 2\rho) }{2(k+y)} - 2 \pi i \frac{c}{24} \right) \delta_{ab} ~,
\label{T-matrix-unrefined}
\end{equation}
where $\Lambda(a)$ is the dominant highest-weight of integrable representation $a$ of algebra $\frakg$, $c$ is the central charge and $y$ is the dual Coxeter number of the algebra (see appendix \ref{appendix-B} for various definitions including $\mathcal{S}$ and $\mathcal{T}$ matrix elements). The quantum state for $T(2,2n)$ links will be given as:
\begin{equation}
\ket{T(2,2n)} = \sum_{a,b} C_{ab}  \ket{a,b} ~.
\label{T(2,2n)-link-state}
\end{equation}
The reduced density matrix can be obtained by tracing out one of the components and its elements, labeled by the representations, are given as:
\begin{equation}
\rho_{ab} = \dfrac{\sum_{m} \left( \left|\dfrac{\left(\mathcal{S}\mathcal{T}^n \mathcal{S}\right)_{0m}}{\mathcal{S}_{0m}}\right|^2 \mathcal{S}_{am} \mathcal{S}_{bm}^{*} \right)}{\sum_{m}\left(\left|\dfrac{\left(\mathcal{S} \mathcal{T}^n \mathcal{S}\right)_{0m}}{\mathcal{S}_{0m}}\right|^2\right)} ~,
\label{T(2,2n)-density-matrix}
\end{equation}
where $a$ and $b$ are the integrable representations and label the $ab$ element of $\rho$ matrix. The link state $\ket{T(2,2n)}$ shows a periodic entanglement structure as we increase the number of twists $n$. We will discuss this feature numerically in the following sections.
 
\subsection{Periodic entanglement structure for $S^3 \backslash T(2,2n)$}
The matrix $\mathcal{T}$ given in eqn.(\ref{T-matrix-unrefined}) is a diagonal matrix and the elements of $\mathcal{T}^n$ are given as: 
\begin{equation}
\mathcal{T}_{aa}^{\,n} = \text{exp}\left( 2 \pi i n \, \frac{\Lambda(a) \bigcdot (\Lambda(a) + 2\rho) }{2(k+y)} - 2 \pi i n \frac{c}{24} \right) ~.
\end{equation}
For any algebra of a given rank $r$ and level $k$, there exists a minimum value of $n$, say $n=n_0(k,r)$, for which we will have:
\begin{equation}
\mathcal{T}^{\,n_0} = \text{exp}\left(- \frac{\pi i n_0 c}{12} \right) \text{diag}(1,1,\ldots,1) ~.
\end{equation}
Thus, the $\mathcal{T}^{\,n_0}$ matrix becomes proportional to identity matrix. We have given the values of $n_0$ for the classical affine Lie algebras in Appendix \ref{appendix-B} (see eqns. (\ref{n0-for-AN}), (\ref{n0-for-BN}), (\ref{n0-for-CN}) and (\ref{n0-for-DN})). At this value of $n=n_0$, the $T(2,2n_0)$ link state in eqn.(\ref{T(2,2n)-link-state}) becomes a direct product state\footnote{Notice that for the classical affine Lie algebras, the $\mathcal{S}$ matrix satisfies $(\mathcal{S}^2)_{0m}=0$, when $m \neq 0$. We have used this property here.}:
\begin{equation}
\ket{T(2,2n_0)} \propto \left(\sum_{a} \frac{\mathcal{S}_{a0}}{\sqrt{\mathcal{S}_{00}}} \ket{a} \right) \otimes \left(\sum_{b} \frac{\mathcal{S}_{b0}}{\sqrt{\mathcal{S}_{00}}} \ket{b}\right) ~.
\end{equation}
Hence the R\'enyi entropy corresponding to this link state vanishes. However, we observe that the entanglement spectrum shows a periodic behavior with fundamental period $\textsf{p}(k,r)$, which is not necessarily $n_0$. In fact, for SU($N$) and Sp($2N$) groups, they seem to be simply related:
\begin{align}
 \text{SU}(N)_{k \geq 1}: \textsf{p} = \begin{cases}
   n_0/2,  & \text{for $N=$ even} \\
	 n_0,  & \text{for $N=$ odd}
	\end{cases} \quad;\quad
	\text{Sp}(2N)_{k \geq 1}: \quad \textsf{p} = n_0/2 
  ~.
	\label{Relation-p-with-n0}
\end{align}
Thus the eigenvalues of reduced density matrix (\ref{T(2,2n)-density-matrix}) and hence the R\'enyi entropy shows the periodicity:
\begin{equation}
\boxed{\lambda \left( n + \textsf{p}\mathbb{Z} \right) = \lambda \left( n \right) \implies \mathcal{R}_{\alpha} \left( n + \textsf{p}\mathbb{Z} \right) = \mathcal{R}_{\alpha} \left( n \right)} ~.
\label{property-entanglement-2}
\end{equation}
The R\'enyi entropy vanishes at $n = \textsf{p}$ and also has a symmetry within the fundamental period in the sense that the R\'enyi entropies for $n$ twists and $(\textsf{p}-n)$ twists ($n \leq \textsf{p}$) are the same (for a given group $G$ and level $k$):
\begin{equation}
\boxed{\mathcal{R}_{\alpha} \left(\textsf{p} \right) = 0 \quad;\quad \mathcal{R}_{\alpha} \left(n \right) = \mathcal{R}_{\alpha} \left( \textsf{p}-n \right), \forall \, n \leq \textsf{p}} ~.
\label{property-entanglement}
\end{equation}
We have tabulated the values of von Neumann entropies ($\mathcal{R}_{1}$) for some of the groups at certain levels in Table-\ref{EE-table}, which shows the periodic structure of $T(2,2n)$ link state, and also confirms eqn.(\ref{property-entanglement}). 
\begin{table}
\begin{adjustwidth}{-1.65cm}{-1cm}
	\centering
	\resizebox{\columnwidth}{!}{
		\begin{tabular}{|c|ccccc|}
		\hline
				\backslashbox{$\boldsymbol{n}$}{$\boldsymbol{G}$ }& $ \boldsymbol{\textbf{SU}(2)_3}$ & $\boldsymbol{\textbf{SU}(3)_2}$ & $\boldsymbol{\textbf{Sp}(6)_2}$ & $\boldsymbol{\textbf{SO}(8)_2}$ & $\boldsymbol{\textbf{SO}(9)_1}$ \\ \hline
		$\boldsymbol{1}$ & $\ln 4$ & $\ln 6$ & $\ln 10$ &  $\ln 11$ & $\ln 3$ \\ 
		$\boldsymbol{2}$ & $\ln \left(3 \left(\frac{3-\sqrt{5}}{2}\right)^{\frac{\sqrt{5}}{3}}\right)$ & $\ln \left(9 \left(\frac{7-3\sqrt{5}}{2}\right)^{\frac{\sqrt{5}}{6}}\right)$ & $\frac{2}{3 \sqrt{3}}\ln \left(\frac{97-56 \sqrt{3}}{\left(54 \sqrt{2}\right)^{-\sqrt{3}}} \right)$ & $\ln \left( \frac{3^{18/13}}{13} \right)$ & $\sqrt{2}\ln \left( \frac{3-2 \sqrt{2}}{2^{\sqrt{2}}} \right)$ \\ 
		$\boldsymbol{3}$ & $\ln \left(6 \left(\frac{7-3\sqrt{5}}{2}\right)^{\frac{\sqrt{5}}{6}}\right)$ & $\ln \left(3 \left(\frac{7-3\sqrt{5}}{2}\right)^{\frac{\sqrt{5}}{6}}\right)$ & $\ln \left( \frac{28 \left(2-\sqrt{3}\right)^{\frac{3 \sqrt{3}}{7}}}{3^{27/28}} \right)$ &  $\ln 11$ & $\ln 3$  \\ 
		$\boldsymbol{4}$ & $\ln 2$ & $\ln 6$ & $\ln \left( \frac{9 \left(2-\sqrt{3}\right)^{\frac{4}{3 \sqrt{3}}}}{2^{8/9}} \right)$ & $\ln \left( \frac{5^{5/8}}{8}\right)$ & $\ln 2$ \\ 
		$\boldsymbol{5}$ & $\ln 2$ & $\ln 3$ & $\ln \left( 34 \left(7-4 \sqrt{3}\right)^{\frac{8 \sqrt{3}}{17}} \right)$ & $\ln 11$ & $\ln 3$ \\ 
		$\boldsymbol{6}$ & $\ln 2$ & $\ln 2$ & $\ln \left( 8 \left(2-\sqrt{3}\right)^{\frac{\sqrt{3}}{2}} \right)$ & $\ln \left( \frac{3^{18/13}}{13} \right)$ & $\sqrt{2}\ln \left( \frac{3-2 \sqrt{2}}{2^{\sqrt{2}}} \right)$ \\ 
		$\boldsymbol{7}$ & $\ln \left(6 \left(\frac{7-3\sqrt{5}}{2}\right)^{\frac{\sqrt{5}}{6}}\right)$ & $\ln \left(9 \left(\frac{7-3\sqrt{5}}{2}\right)^{\frac{\sqrt{5}}{6}}\right)$ & $\ln \left( 34 \left(7-4 \sqrt{3}\right)^{\frac{8 \sqrt{3}}{17}} \right)$ & $\ln 11$ & $\ln 3$ \\ 
		$\boldsymbol{8}$ & $\ln \left(3 \left(\frac{3-\sqrt{5}}{2}\right)^{\frac{\sqrt{5}}{3}}\right)$ & $\ln \left(9 \left(\frac{7-3\sqrt{5}}{2}\right)^{\frac{\sqrt{5}}{6}}\right)$ & $\ln \left( \frac{9 \left(2-\sqrt{3}\right)^{\frac{4}{3 \sqrt{3}}}}{2^{8/9}} \right)$ & $\ln \left( \frac{5^{5/8}}{8}\right)$ & $\boldsymbol{0}$ \\ 
		$\boldsymbol{9}$ & $\ln 4$ & $\ln 2$ & $\ln \left( \frac{28 \left(2-\sqrt{3}\right)^{\frac{3 \sqrt{3}}{7}}}{3^{27/28}} \right)$ & $\ln 11$ &  $\ln 3$ \\ 
		$\boldsymbol{10}$ & $\boldsymbol{0}$ & $\ln 3$ & $\frac{2}{3 \sqrt{3}}\ln \left(\frac{97-56 \sqrt{3}}{\left(54 \sqrt{2}\right)^{-\sqrt{3}}} \right)$ & $\ln \left( \frac{3^{18/13}}{13} \right)$ & $\sqrt{2}\ln \left( \frac{3-2 \sqrt{2}}{2^{\sqrt{2}}} \right)$ \\ 
		$\boldsymbol{11}$ & $\ln 4$ & $\ln 6$ & $\ln 10$ & $\ln 11$ & $\ln 3$ \\ 
		$\boldsymbol{12}$ & $\ln \left(3 \left(\frac{3-\sqrt{5}}{2}\right)^{\frac{\sqrt{5}}{3}}\right)$  & $\ln \left(3 \left(\frac{7-3\sqrt{5}}{2}\right)^{\frac{\sqrt{5}}{6}}\right)$ & $\boldsymbol{0}$ & $\ln \left( \frac{5^{5/8}}{8}\right)$ & $\ln 2$ \\
		$\boldsymbol{13}$ & $\ln \left(6 \left(\frac{7-3\sqrt{5}}{2}\right)^{\frac{\sqrt{5}}{6}}\right)$ & $\ln \left(9 \left(\frac{7-3\sqrt{5}}{2}\right)^{\frac{\sqrt{5}}{6}}\right)$ & $\ln 10$ & $\ln 11$ & $\ln 3$ \\ 
		$\boldsymbol{14}$ & $\ln 2$ & $\ln 6$ & $\frac{2}{3 \sqrt{3}}\ln \left(\frac{97-56 \sqrt{3}}{\left(54 \sqrt{2}\right)^{-\sqrt{3}}} \right)$ & $\ln \left( \frac{3^{18/13}}{13} \right)$ & $\sqrt{2}\ln \left( \frac{3-2 \sqrt{2}}{2^{\sqrt{2}}} \right)$ \\ 
		$\boldsymbol{15}$ & $\ln 2$ & $\boldsymbol{0}$  & $\ln \left( \frac{28 \left(2-\sqrt{3}\right)^{\frac{3 \sqrt{3}}{7}}}{3^{27/28}} \right)$ & $\ln 11$ & $\ln 3$ \\ 
		$\boldsymbol{16}$ & $\ln 2$ & $\ln 6$  & $\ln \left( \frac{9 \left(2-\sqrt{3}\right)^{\frac{4}{3 \sqrt{3}}}}{2^{8/9}} \right)$ & $\boldsymbol{0}$ & $\boldsymbol{0}$ \\ \hline
		\end{tabular} }
		\end{adjustwidth}
		\caption{Table listing the entanglement entropy $\mathcal{R}_1$ for $T(2,2n)$ link state, for some of the groups $G$ and level $k$. The values are given for $1\leq n \leq 16$. We can see that within a group and level, we have a periodic behavior with fundamental period $\textsf{p}$ given in eqn.(\ref{fundamental-period}). We can also see that $\mathcal{R}_1(\textsf{p})=0$ and $\mathcal{R}_1(n)=\mathcal{R}_1(\textsf{p}-n), \forall \, n \leq \textsf{p}$.}
		\label{EE-table}
\end{table}
We find the following values of fundamental period \textsf{p} for various groups:
\begin{align}
 & \boxed{\text{SU}(N)}: \quad  \textsf{p} =
{\begin{cases}
    N(k+N), & \text{for $k>1$} \\
    N,  & \text{for $k=1$}
\end{cases}} \nonumber \\
& \boxed{\text{Sp}(2N)}: \quad  \textsf{p} = {\begin{cases}
    2, & \text{for Sp(2) with $k=1$,} \\
    2(k+N+1),  & \text{otherwise}
\end{cases}} \nonumber \\
& \boxed{\text{SO}(N)}: \quad  
{\begin{cases}
k>1: \quad
  \textsf{p} =\begin{cases}
	  2(k+2), & \text{for SO(3)} \\
    2(k+N-2), & \text{for $N=0,1$ (mod 4)} \\
   4(k+N-2), & \text{for $N=2,3$ (mod 4)}
	\end{cases} \\[25pt]
k=1: \quad
  \textsf{p} =\begin{cases}
	 2, & \text{for SO(3)} \\
   2, & \text{for $N=0$ (mod 4)} \\
   8, & \text{for $N=1,3$ (mod 4)} \\
	 4, & \text{for $N=2$ (mod 4)}
	\end{cases}
\end{cases}} ~.
\label{fundamental-period}
\end{align}

So, we see that the R\'enyi entropy for the link state $\ket{T(2, 2\textsf{p}\mathbb{Z})}$, computed in the Chern-Simons theory for a given group $G$ at a given level $k$ vanishes, though the two boundaries of $S^3 \backslash T(2, 2\textsf{p}\mathbb{Z})$ have a non-trivial topological linking. Ideally, we could find a new invariant for these links such that the R\'enyi entropy is non-vanishing for all values of $n$ for a given $k$ and rank $r$. Indeed we can achieve such a non-zero R\'enyi entropy for all $T(2,2n)$ links using refined link invariants within refined Chern-Simons theory, which we elaborate in the following subsection.

\subsection{Refined entanglement structure of $S^3 \backslash T(2,2n)$}
We can refine the entanglement structure by using the refined Chern-Simons invariants for $T(2,2n)$ links, which will be given as:
\begin{equation}
C_{ab} =  \sum_{m} \frac{\left(\mathcal{S}^{\textsf{ref}}(\mathcal{T}^{\textsf{ref}})^n \mathcal{S}^{\textsf{ref}}\right)_{0m}}{\mathcal{S}^{\textsf{ref}}_{0m}} \,\, \mathcal{S}^{\textsf{ref}}_{am} \, \mathcal{S}^{\textsf{ref}}_{bm} ~,
\label{n-twist-invariant-refined}
\end{equation}
where $\mathcal{S}^{\textsf{ref}}$ and $\mathcal{T}^{\textsf{ref}}$ are the refined versions of ordinary $\mathcal{S}$ and $\mathcal{T}$ matrices respectively. The refinement of the Chern-Simons theory for SU($N$) group was obtained in \cite{Aganagic:2011sg} where the Schur polynomials (which are the characters of integrable representations\footnote{See Appendix \ref{appendix-B} and \cite{fulton2013representation} for more details.}) are lifted to Macdonald polynomials. These Macdonald polynomials have an extra refinement parameter $\beta \in \mathbb{R}$ and in the limit of $\beta \to 1$, we will get back the unrefined version. In the refined case, the elements of modular matrix $\mathcal{T}$ transformation are given as:
\begin{align}
\mathcal{T}^{\textsf{ref}}_{ab} = \text{exp}\left( 2 \pi i\, \frac{\Lambda(a) \bigcdot (\Lambda(a) + 2\beta \rho) }{2(k+\beta y)} + \ldots \right) \delta_{ab} ~,
\label{refined-T}
\end{align}
where we have omitted the terms which are independent of representation and will not affect the computation of R\'enyi entropy. Note that the refinement of the Chern-Simons theory does not change the number of integrable representations and will still be given as $\mathcal{I}_{\frakg}$. In eqn.(\ref{refined-T}), $a$ and $b$ are the integrable representations and $\Lambda(a)$ is the dominant highest weight of representation $a$. Writing this highest weight in terms of Dynkin labels as $\Lambda(a) = a_1 \Lambda_1 + \ldots + a_r \Lambda_r$, we can write the following:
\begin{equation}
\left(\mathcal{T}^{\textsf{ref}}_{aa} \right)^{n} = \text{exp}\left( 2 \pi i n \sum_{i,j=1}^{r} F_{ij}  \frac{a_i(a_j+2 \beta)}{2(k+\beta y)} \right) ~.
\label{refined-T-power-n}
\end{equation}
The first diagonal element corresponding to trivial representation is 1, i.e. $\mathcal{T}^{\textsf{ref}}_{00}=1$. In order to get $(\mathcal{T}^{\textsf{ref}})^{n} \propto \mathbb{1}$, for some value of $n$, all the elements of $(\mathcal{T}^{\textsf{ref}})^{n}$ has to be 1. Of course, we can set $\beta \in \mathbb{R}$ to such a value that this will never happen. As an example, consider SU(2) group and $k=3$ for which we will have:
\begin{equation}
\left(\mathcal{T}^{\textsf{ref}} \right)^{n} = \left(
\begin{array}{cccc}
 1 & 0 & 0 & 0 \\
 0 & e^{\frac{i n \pi  (2 \beta +1)}{4 \beta +6}} & 0 & 0 \\
 0 & 0 & e^{\frac{2 i n \pi  (\beta +1)}{2 \beta +3}} & 0 \\
 0 & 0 & 0 & e^{\frac{3 i n \pi }{2}} \\
\end{array}
\right) ~.
\label{refined-SU2-k=3-T-power-n}
\end{equation}
In the unrefined case ($\beta=1$), we will get $\mathcal{T}^{20} = \mathbb{1}$ and the entanglement structure will have a periodicity $\textsf{p} = 10$, which we have given in Table-\ref{EE-table}. However, if we take $\beta$ to be some irrational number, then $(\mathcal{T}^{\textsf{ref}})^{n} \neq \mathbb{1}$, for any integer value of $n$. This indicates that the R\'enyi entropy from refined link invariants can probe richer topological linking information than that of unrefined ones.


\section{Conclusion and discussion}
\label{sec5}
In this note, we studied the entanglement structure corresponding to bi-partitioning of link states in Chern-Simons theory with a generic gauge group. In the large Chern-Simons coupling limit $(k\to\infty)$ or large rank $(r\to\infty)$ limit, the upper bound of R\'enyi entropy goes as $\ln k$ or $\ln r$, which indicates that R\'enyi entropy can not diverge faster than $\ln k$ or $\ln r$. We also considered the limit with both large $k$ and large $r$ but a finite ratio of them.

For a class of torus links $T(2,2n)$, we obtained the eigenvalues of the reduced density matrix numerically for $n>2$ and analytically for $n=2$. In the latter case, we verified that R\'enyi entropy converges in the limit $k\to\infty$ for any $\alpha>0$.
From the entanglement entropy for links $T(2,2n)$ as presented in Table-\ref{EE-table}, we infer its periodic behavior in $n$ for a given rank $r$ and level $k$. We have analytically deduced the periodicity $\mathsf{p}(k,r)$ for various classical Lie groups (eqn.(\ref{fundamental-period})) such that R\'enyi entropy vanishes whenever $n = 0 \text{ (mod } \textsf{p})$.

With the hope of achieving non-vanishing R\'enyi entropy for links $T(2,2n)$ at any finite $n$, we considered refined Chern-Simons theory and their invariants. Interestingly, R\'enyi entropy from $SU(2)$ refined invariants does capture richer topological linking information. 

Another interesting question is whether R\'enyi entropy will be the same for two links which share the same Jones' polynomial. In fact, there are certain class of links with trivial Jones' polynomial \cite{eliahou2003infinite}. We checked some examples of links sharing the same Jones'  polynomial but different colored Jones and observed that the R\'enyi entropies are different.

For a two-component link $\mathcal{L}_1$ obtained by replacing one of the circles of Hopf link by knot $\mathcal{K}_1$ (i.e. $\mathcal{L}_1 = \mathcal{K}_1 \# \mathrm{H}$, where $\mathrm{H}$ denotes the Hopf link and $\#$ is the connected sum) and link $\mathcal{L}_2$ obtained by replacing one of the circles by knot $\mathcal{K}_2$ ($\mathcal{L}_2 = \mathcal{K}_2 \# \mathrm{H}$),  
the $SU(2)$ R\'enyi entropy for the two links will be the same if knots $\mathcal{K}_1$ and $\mathcal{K}_2$ are mutant knots. This is because the mutant knot pairs have the same colored Jones' polynomial. Hence to distinguish such links, we have to go to other gauge groups.  Note that a colored HOMFLY-PT for mixed representation of SU($N$) were shown to be different for a mutant pair $\mathcal{K}_1$, $\mathcal{K}_2$ \cite{morton1996distinguishing, Nawata:2015xha, Mironov:2015aia}. This indicates that probably SU($N$) R\'enyi entropy could distinguish such links. However,  at the moment we do not have a complete solution of SU($N$) link invariants in variable $q$ for arbitrary representations.

It is always interesting to generalize the current discussion to multi-component links with generic gauge group. Further, just like our analysis for $T(2,2n)$ links, it would be worthwhile to explore other torus and non-torus links to see if the R\'enyi entropy has a periodic behavior. We hope to report progress in future work.

\appendix
\smallsection
\section{Integrable representations of affine Lie algebra at level $k$}
\label{appendix-A}
\small{
In this appendix, we briefly review on how to obtain the integrable representations for the affine Lie algebra $\hat{g}$ at the level $k$. We will use the following notations. Consider a Lie group $G$ with the associated Lie algebra $\frakg$ which has rank $r$ and dimension $d$. The Lie algebra has $d-r$ roots and any root can be expressed in terms of $r$ linearly independent roots, called the simple roots $\alpha_i$ of $\frakg$. The simple coroots which we denote as $\beta_i$ are defined as,
\begin{equation}
\beta_i \equiv \frac{2 \alpha_i}{\alpha_i \bigcdot \alpha_i} \quad;\quad 1 \leq i \leq r ~,
\end{equation}
where $\alpha_i \bigcdot \alpha_i$ denotes the length of the root $\alpha_i$. A Lie algebra has a special root $\theta$ called the highest root of $\frakg$ and can be expanded in terms of simple roots or coroots as:
\begin{equation}
\theta = \sum_{i=1}^{r} \theta_i \alpha_{i} = \sum_{i=1}^{r} \phi_i \beta_{i} ~.
\end{equation}
The integers $\theta_i$ and $\phi_i$ are non-negative and are called the Kac labels and dual Kac labels of the Lie algebra $\frakg$ (also called as marks and comarks respectively). The fundamental weights of the Lie algebra $\frakg$ are $\Lambda_i$ and any weight $\lambda$ can be expanded in the basis of fundamental weights as:
\begin{equation}
\lambda = \sum_{i=1}^{r} \lambda_i \Lambda_{i} ~.
\end{equation}
The expansion coefficients $\lambda_i$ are integers and are called the Dynkin labels of weight $\lambda$. A weight $\lambda$ belongs to the fundamental chamber of the Weyl group if all its Dynkin labels are non-negative integers ($\lambda_i \geq 0$). Such weights are also called as dominant weights. The representation $a$ of the Lie algebra $\frakg$ has an associated dominant highest-weight $\Lambda(a)$ with Dynkin labels $a_1, a_2, \ldots, a_r$. We will use these Dynkin labels to specify the representation: $a = \left[a_1, a_2, \ldots a_r \right]$.

Any finite Lie algebra $\frakg$ can be extended to an affine Lie algebra $\hat{\frakg}$ by adding an extra node to the Dynkin diagram of $\frakg$. As a result, the weight $\hat{\lambda}$ of the affine Lie algebra $\hat{\frakg}$ carry an extra Dynkin label $\lambda_0$. Thus we can denote an affine weight $\hat{\lambda}$ in terms of its Dynkin labels as: $\hat{\lambda} = \left[\lambda_0, \lambda_1, \lambda_2, \ldots \lambda_r \right]$, where $\lambda_1, \ldots \lambda_r$ are the usual Dynkin labels of the weight $\lambda$ of Lie algebra $\frakg$. The zeroth Dynkin label $\lambda_0$ is related to the other Dynkin labels and the level $k$ as the following:
\begin{equation}
\lambda_0 = k - \sum_{i=1}^{r} \phi_i \lambda_{i} ~,
\end{equation}
where $\phi_i$ are the dual Kac labels of algebra $\frakg$ defined earlier. The dominant weights are the ones which belong to the fundamental chamber of the affine Weyl group and have all the Dynkin labels as non-negative integers ($\lambda_i \geq 0$ for $0 \leq i \leq r$). A representation $\hat{a}$ of the affine Lie algebra is integrable if it has a dominant highest weight $\hat{\Lambda}(a)$, i.e. if $a_i \geq 0$ for $i = 0, 1, \ldots, r$. Thus the integrability condition is: 
\begin{equation}
a_0 = k - \sum_{i=1}^{r} \phi_i a_{i} \quad;\quad a_0, a_1, \ldots, a_r \geq 0, \quad k \geq 0 ~.
\label{Dynkin-labels}
\end{equation}
Since the zeroth Dynkin label $a_0$ depends upon the level $k$ and other Dynkin labels, one can restrict the representations of the Lie algebra $\frakg$ for a finite value of $k$. The Dynkin labels of an integrable representation $a = \left[a_1, a_2, \ldots a_r \right]$ of $\frakg$ will satisfy (using $a_0 \geq 0$),
\begin{equation}
\sum_{i=1}^{r} \phi_i a_{i} \leq k \quad;\quad a_i \geq 0 ~.
\label{Integrability}
\end{equation}

\subsection*{\small{Generating function for number of integrable representations}}
For a given $k$, one can count the number of integrable representations $\mathcal{I}_{\frakg}$ of algebra $\frakg_k$ as following:
\begin{align}
\boxed{\mathcal{I}_{\frakg} = \sum_{m=0}^k \mathcal{I}_m \quad;\quad \sum_{n=0}^{\infty} \mathcal{I}_n x^n = \prod_{i=1}^r \left( \dfrac{1}{1-x^{\phi_i}} \right) = \sum_{m_1, \ldots, m_r = 0}^{\infty} x^{m_1 \phi_1 + \ldots + m_r \phi_r}  } ~,
\label{Integrable-rep-arbit-algebra}
\end{align}
where the coefficient $\mathcal{I}_n$ of the generating series counts the number of possible solutions of $\sum_{i=1}^{r} \phi_i a_{i} = n$.

\subsection*{\small{For classical Lie algebra $\boldsymbol{A_{N-1}=\mathfrak{su}(N)}$}}
The highest root and hence the dual Kac labels are given as:
\begin{equation}
\theta = \sum_{i=1}^{N-1} \beta_{i} \implies \phi_i = 1 ~.
\end{equation}
Thus the integrability condition of eqn.(\ref{Integrability}) becomes, 
\begin{equation}
a_{1} + a_2 + \ldots + a_{N-1} \leq k \quad;\quad a_i \geq 0 ~.
\label{Integrability-SUN}
\end{equation}
Counting the number of integrable representations for a given value of level $k$ and $N$, one obtains:
\begin{equation}
\mathcal{I}_{\mathfrak{su}(N)} = \binom{k+N-1}{N-1} = \dfrac{(k+N-1)!}{(N-1)!\,k!} ~.
\label{Number-Integrability-SUN}
\end{equation}

\subsection*{\small{For classical Lie algebra $\boldsymbol{B_{N}=\mathfrak{so}(2N+1)}$}}
The dual Kac labels are:
\begin{equation}
\phi_1 = \phi_N = 1 \quad,\quad \phi_2 = \ldots = \phi_{N-1} = 2 ~.
\end{equation}
The integrability condition of eqn.(\ref{Integrability}) becomes\footnote{See the integrability condition given in \cite{Mlawer:1990uv} for $\mathfrak{so}(3)$ algebra, where the symbol $K$ (used to denote $\text{SO}(3)_{K}$) is related to the actual level $k$ as $K=\frac{k}{2}$.}, 
\begin{align}
& N=1: a_1 \leq k  \quad;\quad N=2:a_1+a_2 \leq k, \nonumber \\
& N \geq 3: a_{1} + 2\,a_2 + 2\,a_3 + \ldots + 2\,a_{N-1} + a_N \leq k ~.
\label{Integrability-BN}
\end{align}
Counting the number of integrable representations for a given value of level $k$ and $N$:
\begin{equation}
\mathcal{I}_{\mathfrak{so}(2N+1)} = 
{\begin{cases}
    k+1 & \text{for $\mathfrak{so}(3)$} \\[10pt]
    \dfrac{\left(\frac{k}{2}+N-1 \right)!}{N!\,\left(\frac{k}{2} \right)!}(2k+N), & \text{for even $k$} \\[15pt]
    \dfrac{\left(\frac{k-1}{2}+N-1 \right)!}{N!\,\left(\frac{k-1}{2} \right)!}(2k-2+3N), & \text{for odd $k$}
\end{cases}}   ~.
\label{Number-Integrability-BN}
\end{equation}

\subsection*{\small{For classical Lie algebra $\boldsymbol{C_{N}=\mathfrak{sp}(2N)}$}}
All the dual Kac labels are equal to one and hence the integrability condition is, 
\begin{equation}
a_{1} + a_2 + a_3 + \ldots + a_N \leq k ~.
\label{Integrability-CN}
\end{equation}
This is exactly the same condition as that of $A_N$ algebra and the counting gives:
\begin{equation}
\mathcal{I}_{\mathfrak{sp}(2N)} = \dfrac{(k+N)!}{N!\,k!} ~.
\label{Number-Integrability-CN}
\end{equation}

\subsection*{\small{For classical Lie algebra $\boldsymbol{D_{N}=\mathfrak{so}(2N)}$}}
The dual Kac labels for $\mathfrak{so}(2N)$ are:
\begin{equation}
\phi_1 = \phi_{N-1} = \phi_N = 1 \quad,\quad \phi_2 = \ldots = \phi_{N-2} = 2 ~.
\end{equation}
The integrability condition becomes, 
\begin{align}
& N=2: a_1 \leq k; a_2 \leq k \quad;\quad N=3:a_1+a_2+a_3 \leq k, \nonumber \\
& N \geq 4: a_{1} + 2\,a_2  + \ldots + 2\,a_{N-2} + a_{N-1} + a_N \leq k ~.
\label{Integrability-DN}
\end{align}
Counting the number of integrable representations for a given value of level $k$ and $N$ gives:
\begin{equation}
\mathcal{I}_{\mathfrak{so}(2N)} = 
{\begin{cases}
    \dfrac{\left(\frac{k}{2}+N-2 \right)!}{N!\,\left(\frac{k}{2} \right)!}(2k^2+ (N-1)(4k+N)), & \text{for even $k$} \\[20pt]
    \dfrac{\left(\frac{k-1}{2}+N-1 \right)!}{N!\,\left(\frac{k-1}{2} \right)!}(4k-4+4N), & \text{for odd $k$}
\end{cases}} ~.
\label{Number-Integrability-DN}
\end{equation}

\subsection*{\small{For exceptional Lie algebra $\boldsymbol{E_6}$}}
For Lie algebra $E_6$, the dual Kac labels are:
\begin{equation}
\phi_1 = 1,\,\, \phi_2 = 2,\,\, \phi_3 = 3,\,\, \phi_4 = 2,\,\, \phi_5 = 1,\,\, \phi_6 = 2 ~.
\end{equation}
The integrability condition of eqn.(\ref{Integrability}) becomes, 
\begin{equation}
a_{1} + 2\,a_2 + 3\,a_3 + 2\,a_4 + a_5 + 2\,a_6 \leq k \quad;\quad a_i \geq 0 ~.
\label{Integrability-E6}
\end{equation}
The number of integrable representations for a given value of level $k$ is given as:
\begin{equation}
\mathcal{I}_{E_6} = 
{\begin{dcases}
    \dfrac{(k+6)^2 (k^4 + 24k^3 + 186k^2 + 504k + 480)}{17280}, & \text{for $k=0$ (mod 6);} \\[10pt]
    \dfrac{(k+1)(k+4)(k+5)(k+7)(k+8)(k+11)}{17280}, & \text{for $k=1, 5$ (mod 6);} \\[10pt]
		\dfrac{(k+2)(k+4)(k+8)(k+10)(k^2+12k+26)}{17280}, & \text{for $k=2, 4$ (mod 6);} \\[10pt]
		\dfrac{(k+3)(k+9) (k^4 + 24k^3 + 195k^2 + 612k + 480)}{17280}, & \text{for $k=3$ (mod 6).}
\end{dcases}}
\label{Number-Integrability-E6}
\end{equation}

\subsection*{\small{For exceptional Lie algebra $\boldsymbol{E_7}$}}
For Lie algebra $E_7$, the dual Kac labels are:
\begin{equation}
\phi_1 = 2,\,\, \phi_2 = 3,\,\, \phi_3 = 4,\,\, \phi_4 = 3,\,\, \phi_5 = 2,\,\, \phi_6 = 1,\,\, \phi_7 = 2 ~.
\end{equation}
The integrability condition of eqn.(\ref{Integrability}) becomes, 
\begin{equation}
2\,a_{1} + 3\,a_2 + 4\,a_3 + 3\,a_4 + 2\,a_5 + a_6 + 2\,a_7 \leq k \quad;\quad a_i \geq 0 ~.
\label{Integrability-E7}
\end{equation}
The number of integrable representations for a given value of level $k$ is given as:
\begin{equation}
\resizebox{\textwidth}{!}{
$ \mathcal{I}_{E_7} = 
\begin{dcases}
    \frac{(k+12) \left(k^6+51 k^5+1005 k^4+9675 k^3+47784 k^2+116064 k+120960\right)}{1451520} & \, \text{for $k=0$ (mod 12) ;} \\[6pt]
    \frac{(k+1) (k+5) (k+7) (k+9) (k+11) (k+13) (k+17)}{1451520} & \, \text{for $k=1, 5, 7, 11$ (mod 12) ;} \\[6pt]
		\frac{(k+2) (k+10) (k+13) (k+14) \left(k^3+24 k^2+155 k+342\right)}{1451520} & \, \text{for $k=2, 10$ (mod 12) ;} \\[6pt]
		\frac{(k+3) (k+9) (k+15) \left(k^4+36 k^3+438 k^2+2052 k+2289\right)}{1451520} & \, \text{for $k=3, 9$ (mod 12) ;} \\[6pt]
		\frac{(k+4) (k+5) (k+8) (k+16) \left(k^3+30 k^2+263 k+504\right)}{1451520} & \, \text{for $k=4, 8$ (mod 12) ;} \\[6pt]
		\frac{(k+6) \left(k^6+57 k^5+1275 k^4+14085 k^3+79374 k^2+213228 k+234360\right)}{1451520} & \, \text{for $k=6$ (mod 12) .}
\end{dcases}$ }
\label{Number-Integrability-E7}
\end{equation}

\subsection*{\small{For exceptional Lie algebra $\boldsymbol{E_8}$}}
For Lie algebra $E_8$, the dual Kac labels $E_8$ are:
\begin{equation}
\phi_1 = 2,\,\, \phi_2 = 3,\,\, \phi_3 = 4,\,\, \phi_4 = 5,\,\, \phi_5 = 6,\,\, \phi_6 = 4,\,\, \phi_7 = 2,\,\, \phi_8 = 3 ~.
\end{equation}
The integrability condition of eqn.(\ref{Integrability}) becomes, 
\begin{equation}
2\,a_{1} + 3\,a_2 + 4\,a_3 + 5\,a_4 + 6\,a_5 + 4\,a_6 + 2\,a_7 + 3\,a_8 \leq k \quad;\quad a_i \geq 0 ~.
\label{Integrability-E8}
\end{equation}
Using the generating function given in eqn.(\ref{Integrable-rep-arbit-algebra}), the number of integrable representations for a given value of level $k$ is given as:
\begin{align}
\mathcal{I}_{E_8} & = \frac{k (k+30) \left(3 k^6+270 k^5+9960 k^4+192600 k^3+2085849 k^2+12355470 k+37972090\right)}{2090188800} \nonumber \\[4pt]
& + \frac{(k+15)}{256} \cos \left(\frac{\pi  k}{2}\right) + \frac{1}{108} \cos \left(\frac{\pi  k}{3}\right) +\frac{\left(2 k^2+60 k+397\right)}{2916} \cos \left(\frac{2 \pi  k}{3}\right) + \frac{2}{125} \cos \left(\frac{2 \pi  k}{5}\right) \nonumber \\[4pt]
& + \frac{2}{125} \cos \left(\frac{4 \pi  k}{5}\right) + \frac{\left(6 k^4+360 k^3+7476 k^2+62280 k+169441\right)}{884736} \cos (\pi  k)+ \frac{1709433137}{2985984000} ~.            
\label{Number-Integrability-E8}
\end{align}

\subsection*{\small{For exceptional Lie algebra $\boldsymbol{F_4}$}}
For Lie algebra $F_4$, the dual Kac labels are:
\begin{equation}
\phi_1 = 2,\,\, \phi_2 = 3,\,\, \phi_3 = 2,\,\, \phi_4 = 1 ~.
\end{equation}
The integrability condition becomes, 
\begin{equation}
2\,a_{1} + 3\,a_2 + 2\,a_3 + a_4 \leq k \quad;\quad a_i \geq 0 ~.
\label{Integrability-F4}
\end{equation}
The number of integrable representations for a given value of level $k$ is given as:
\begin{equation}
\mathcal{I}_{F_4} = 
\begin{dcases}
    \dfrac{(k+6) (k^3 + 12k^2 + 40k + 48)}{288} \quad, & \text{for $k=0$ (mod 6);} \\[10pt]
    \dfrac{(k+1)(k+5)^2(k+7)}{288} \quad, & \text{for $k=1$ or $5$ (mod 6);} \\[10pt]
		\dfrac{(k+2)(k+4)^2(k+8)}{288} \quad, & \text{for $k=2$ or $4$ (mod 6);} \\[10pt]
		\dfrac{(k+3) (k^3 + 15k^2 + 67k + 69)}{288} \quad, & \text{for $k=3$ (mod 6).}
\end{dcases}
\label{Number-Integrability-F4}
\end{equation}

\subsection*{\small{For exceptional Lie algebra $\boldsymbol{G_2}$}}
For Lie algebra $G_2$, the dual Kac labels are:
\begin{equation}
\phi_1 = 2,\,\, \phi_2 = 1 ~.
\end{equation}
The integrability condition of eqn.(\ref{Integrability}) becomes, 
\begin{equation}
2\,a_{1} + a_2 \leq k \quad;\quad a_i \geq 0 ~.
\label{Integrability-G2}
\end{equation}
The number of integrable representations for a given value of level $k$ is given as:
\begin{equation}
\mathcal{I}_{G_2} = 
\begin{dcases}
    \dfrac{(k+2)(k+2)}{4} \quad, & \text{for even $k$;} \\[10pt]
    \dfrac{(k+1)(k+3)}{4} \quad, & \text{for odd $k$.}
\end{dcases}
\label{Number-Integrability-G2}
\end{equation}
}

\smallsection
\section{Modular $\mathcal{S}$ and $\mathcal{T}$ transformation matrices of affine Lie algebra}
\label{appendix-B}
\small{
The modular group is generated by two modular transformations $\mathcal{S}$ and $\mathcal{T}$. The importance of the modular group lies in the fact that the characters of the dominant highest-weight (or the integrable) representations of an affine Lie algebra $\hat{\frakg}$ at some fixed level $k$ transform into each other under the action of the modular group. The action of the modular group on the characters can be encoded in the matrix elements $\mathcal{S}_{\hat{\Lambda}(a)\hat{\Lambda}(b)}$ and $\mathcal{T}_{\hat{\Lambda}(a)\hat{\Lambda}(b)}$ where $\hat{\Lambda}(a)$ and $\hat{\Lambda}(b)$ are respectively the dominant highest-weights of the integrable representations $\hat{a}$ and $\hat{b}$ of affine Lie algebra $\hat{\frakg}$ at level $k$. From now on, we will simply label these elements as $\mathcal{S}_{\hat{a}\hat{b}}$ and $\mathcal{T}_{\hat{a}\hat{b}}$. Thus we will use the same symbol $\hat{a}$ to denote both the integrable representation as well as their affine dominant highest-weights and it should be understood as the former or the latter depending upon the context. As we have mentioned in appendix-\ref{appendix-A}, an affine integrable dominant highest-weight $\hat{a}$ has an extra Dynkin label $a_0$ as compared to that of its finite analogue $a$ which has $r$ Dynkin labels $a_1, \ldots, a_r$ and these are related via eqn.(\ref{Dynkin-labels}). Thus for a given level $k$, if we just keep track of the zeroth Dynkin label, we can write the modular transformations in terms of the integrable representations $a$ and $b$ of the Lie algebra $\frakg$. The matrix elements of the modular transformations are given as \cite{francesco2012conformal}:
\begin{align}
\mathcal{S}_{ab} &= \frac{i^{\left|\Delta_{+}\right|}}{(k+y)^{\frac{r}{2}}}\left(\frac{\text{vol}(P)}{\text{vol}(Q^{\vee})}\right)^{\frac{1}{2}}\sum_{w \in \mathcal{W}} \epsilon(w)\,\, \text{exp}\left(-2 \pi i\frac{w(a+\rho) \bigcdot (b+\rho)}{k+y}\right) \nonumber \\[10pt]
\mathcal{T}_{ab} &= \text{exp}\left( 2 \pi i\frac{a \bigcdot (a + 2\rho) }{2(k+y)} - 2 \pi i \frac{c}{24} \right) \delta_{ab} \nonumber \\
&= \text{exp}\left(-2 \pi i \, \frac{k \, \text{dim}(\frakg)}{24(k+y)} \right) \text{exp}\left( 2 \pi i \sum_{i,j=1}^{r} \frac{a_i(a_j+2)}{2(k+y)} F_{ij} \right)\delta_{ab} ~.
\label{S-T-Weyl-definition}
\end{align}
Here the weights $a$ and $b$ are given in terms of their Dynkin labels, i.e, $a = a_1\Lambda_1+\ldots + a_r\Lambda_r $ and $b = b_1\Lambda_1+\ldots + b_r\Lambda_r $. The Weyl vector $\rho$ is sum of the fundamental weights and hence has all the Dynkin labels 1: $\rho = \Lambda_1+\ldots + \Lambda_r$. The $\bigcdot$ is simply the dot product between the two weight vectors with $y$ denoting the dual Coxeter number of the Lie algebra $\frakg$ having rank $r$. The central charge of the affine algebra $\hat{\frakg}$ is denoted as $c$ and is given as: $c = \frac{k \, \text{dim}(\frakg)}{k+y}$. The scalar product of the fundamental weights is defined as $\Lambda_i \bigcdot \Lambda_j  = F_{ij}$, where $F_{ij}$ is the element of quadratic form matrix of the algebra\footnote{See appendix 13.A of \cite{francesco2012conformal} for the quadratic form matrices of Lie algebras.}. The symbols $P$ and $Q^{\vee}$ denote the weight and coroot lattices and vol denotes the volume of these lattices. The number of positive roots of algebra $\frakg$ is denoted as $\left|\Delta_{+}\right|$. The summation in the definition of $\mathcal{S}$ matrix element is over all the elements $w$ of the Weyl group $\mathcal{W}$.  The Weyl group of a Lie algebra $\frakg$ is generated by the Weyl reflections with respect to the simple roots denoted as $w_{\alpha_1}, w_{\alpha_2}, \ldots, w_{\alpha_r}$. The action of such a reflection on any weight $a = a_1\Lambda_1+\ldots a_r\Lambda_r \equiv [a_1, \ldots, a_r]$ is given as,
\begin{equation}
w_{\alpha_i}(a) = a - (\beta_i \bigcdot a) \alpha_i = a - 2\frac{\alpha_i \bigcdot a}{\alpha_i \bigcdot \alpha_i} \alpha_i ~.
\label{Weyl-reflection}
\end{equation}
Any element $w \in \mathcal{W}$ can be decomposed in the form $w = w_{\alpha_1}^{n_1} w_{\alpha_2}^{n_2}\ldots w_{\alpha_r}^{n_r}$ where $n_i$ are non-negative integers. The signature of such an element of the Weyl group is defined as,
\begin{equation}
\epsilon(w) = (-1)^{n_1 + n_2 + \ldots + n_r} ~.
\end{equation}

\subsection*{\small{Modular $\mathcal{S}$ matrix as characters of algebra $\frakg$}}
The modular $\mathcal{S}$ transformation can be alternatively given in terms of characters of the representations. The whole content of any representation $a$ of the Lie algebra $\frakg$ can be encoded in its character $\chi_{a}$. A representation $a$ of Lie algebra has an associated representation space $R_a$, the states of which are labeled by weights $\lambda \in R_a$. The highest weight in this space is denoted as $\Lambda$ (or simply $a$ by abuse of notation as mentioned earlier). Any other weight $\lambda \in R_a$ can be obtained by the action of lowering operators of algebra $\frakg$ on the highest weight of $R_a$. Moreover a weight $\lambda$ can occur more than once in $R_a$ and this is called as the multiplicity $\text{mul}(\lambda)$ of the weight $\lambda \in R_a$. Note that the highest weight state is unique which means $\text{mul}(\Lambda)=1$. The character of the representation $a$ is defined as,
\begin{equation}
\chi_{a} = \sum_{\lambda \in R_a} \text{mul}(\lambda) \,\, \text{exp}(\lambda) ~.
\end{equation}
As we have already discussed, any weight $\lambda \in R_a$ is usually given in terms of their Dynkin labels: $\lambda = \lambda_1 \Lambda_1 + \ldots + \lambda_r \Lambda_r \equiv [\lambda_1, \ldots, \lambda_r]$ where $\Lambda_i$ are the fundamental weights of the Lie algebra $\frakg$. Thus the character defined above is a function of $r$ variables $e^{\Lambda_1}, e^{\Lambda_2}, \ldots, e^{\Lambda_r}$. Sometimes these characters are evaluated at arbitrary points $e^{x_1}, e^{x_2}, \ldots, e^{x_r}$ and in that case the characters will be denoted as $\chi_{a}[x]$, where $x \equiv [x_1, \ldots, x_r] = x_1 \Lambda_1 + \ldots + x_r \Lambda_r$. The modular $\mathcal{S}$ matrix elements for two representations $a$ and $b$ of Lie algebra $\frakg$ can be given as a function of characters of $a$ and $b$ evaluated at special points:
\begin{align}
\frac{\mathcal{S}_{ab}}{\mathcal{S}_{00}} &=  \chi_{a}\left[\frac{2 \pi i}{k+y} \left(b+\rho \right) \right] \chi_{b}\left[\frac{2 \pi i}{k+y} \, \rho \right] \quad; \nonumber \\ 
\mathcal{S}_{00} &= \frac{1}{(k+y)^{\frac{r}{2}}} \left(\frac{\text{vol}(P)}{\text{vol}(Q^{\vee})}\right)^{\frac{1}{2}} \prod_{\alpha >0} 2 \sin\left(\pi \frac{\alpha \bigcdot \rho}{k+y}\right) ~.
\label{S-matrix-character}
\end{align}
The product in $\mathcal{S}_{00}$ is over all positive roots of algebra $\frakg$. On the right side $a$ and $b$ denote the highest weights of representations $a$ and $b$. The characters of a representation of the Lie algebra can be written in terms of Schur functions and this helps in obtaining an explicit form of the characters. In the following subsections we give the characters of classical Lie algebras in terms of Schur functions and give an explicit expression for the $\mathcal{S}$ and $\mathcal{T}$ matrix elements (see \cite{Mlawer:1990uv} for modular $\mathcal{S}$ transformations of classical Lie groups).

\subsection*{\small{For classical Lie algebra $\boldsymbol{A_{N-1}=\mathfrak{su}(N)}$}}
The character of a representation $a$ of $\mathfrak{su}(N)$ where $a$ is labeled by its highest weight $a=[a_1, \ldots, a_{N-1}]$ and the character being evaluated at variables $e^{x_1}, e^{x_2}, \ldots, e^{x_{N}}$ is given as,
\begin{equation}
\chi_{a}[x] =  \frac{\left| \text{exp} \left[ x_j(l_i + N - i) \right] \right|}{\left| \text{exp} \left[ x_j(N - i) \right] \right|} \quad;\quad l_i \equiv \sum_{m=i}^{N-1}a_m ~,
\end{equation}
where $\left|** \right|$ is the determinant of the matrix whose rows and columns are indexed by $i$ and $j$ with $i,j = 1, 2, \ldots, N$ (note that for $\mathfrak{su}(N)$ algebra, $a_N = 0$ and $x_1 + \ldots + x_N = 0$). Using this definition of character, the explicit expression of $\mathcal{S}$ matrix element for representations $a=[a_1, \ldots, a_{N-1}]$ and $b=[b_1, \ldots, b_{N-1}]$ can be evaluated from eqn.(\ref{S-matrix-character}) and is given as: 
\begin{equation}
\boxed{\mathcal{S}_{ab} = (-i)^{\frac{N(N-1)}{2}} \frac{N^{-\frac{1}{2}}}{(k+N)^{\frac{N-1}{2}}} \left| \text{exp}\left(2 \pi i \frac{f_i(a) f_j(b)}{k+N} \right) \right|} ~,
\end{equation} 
where the quantity $f_i$ for an integrable representation $a$ is defined as:
\begin{equation}
f_i(a) \equiv l_i - i - \frac{1}{N}\left( \sum_{m=1}^{N-1} l_m \right) + \frac{N+1}{2} ~.
\end{equation}
In order to write the elements of $\mathcal{T}$ matrix, we need the elements of quadratic form which is given for $\mathfrak{su}(N)$ as:
\begin{equation}
F_{ij} =  
\begin{dcases}
    \frac{j(N-i)}{N} \quad, & j=1,2,\ldots,i \\[5pt]
    \frac{i(N-j)}{N} \quad, & j=i+1,\ldots,N-1
\end{dcases}, \quad i=1,2,\ldots,N-1 ~.
\end{equation}
Using this, we get the $\mathcal{T}$ matrix elements as:
\begin{equation}
\boxed{\mathcal{T}_{ab} = \text{exp}\left(-2 \pi i \, \frac{k(N^2-1)}{24(k+N)} \right) \text{exp}\left[ \frac{2 \pi i}{2(k+N)} \left(\frac{x_a(N^2-x_a)}{N} + y_a\right) \right]\delta_{ab}} ~,
\end{equation}
where $x_a$ and $y_a$ are integers. For a representation $a=(a_1,\ldots,a_N)$, these integers can be given in terms of Dynkin labels as:
\begin{align}
x_a = \sum_{i=1}^{N-1} ia_i  \quad;\quad y_a = \sum_{i=1}^{N-1} a_i \left(- i^2 + \sum_{j=1}^i ja_j + \sum_{j=i+1}^{N-1} i a_j \right) ~.
\end{align}
Now we are interested in finding a minimum positive integer $n_0$ such that all the elements of matrix $\mathcal{T}^{n_0}$ become equal. In other words, we want $\mathcal{T}^{n_0}$ to be a scalar matrix (proportional to identity matrix). Since $\mathcal{T}$ is a diagonal matrix, we can write:
\begin{equation}
\mathcal{T}_{ab}^{n_0} = \text{exp}\left(-2 \pi i n_0 \, \frac{k(N^2-1)}{24(k+N)} \right) \text{exp}\left[ \frac{2 \pi i n_0}{2(k+N)} \left(\frac{x_a(N^2-x_a)}{N} + y_a\right) \right]\delta_{ab} ~.
\end{equation}
The integer $x_a$ can be either even or odd, but the quantity $x_a(N^2-x_a)$ is always (for all representations) even whenever $N$ is odd. For even values of $N$, this quantity can be either even or odd. Thus for $k\geq 2$, the required $n_0$ will be $2N(k+N)$ or $N(k+N)$ depending on whether $N$ is even or odd, respectively. For $k=1$, the $\mathcal{T}$ matrix elements can be explicitly worked out as:
\begin{equation}
\mathcal{T}_{00} = \text{exp}\left(\frac{\pi i(1-N)}{12} \right) \quad;\quad \mathcal{T}_{rs} = \text{exp}\left(\frac{\pi i(1-N)}{12} \right) \text{exp}\left(\frac{2\pi is(N-s)}{2N} \right)\delta_{rs} ~,
\end{equation}
where $r$ and $s$ denote the $r$-th row and $s$-th column respectively ($r,s=1,\ldots,N-1$). We notice again that $s(N-s)$ is always even for odd values of $N$. Thus the minimum positive integer $n_0$ for affine $\mathfrak{su}(N)$ at level $k$ such that $\mathcal{T}^{n_0}$ is a scalar matrix can be given as:
\begin{equation}
\boxed{\mathfrak{su}(N)_k}: \quad n_0 = \begin{cases}
  2N,  & \text{for $k=1$, even $N$} \\ 
	N,  & \text{for $k=1$, odd $N$} \\
	2N(k+N), & \text{for $k>1$, even $N$} \\
	N(k+N), & \text{for $k>1$, odd $N$}
	\end{cases}  
  ~.
	\label{n0-for-AN}
\end{equation}

\subsection*{\small{For classical Lie algebra $\boldsymbol{B_{N}=\mathfrak{so}(2N+1)}$}}
The character of a representation $a=[a_1, \ldots, a_{N}]$ of $\mathfrak{so}(2N+1)$ evaluated at variables $e^{x_1}, e^{x_2}, \ldots, e^{x_{N}}$ is given as,
\begin{equation}
\chi_{a}[x] = \frac{\left| \text{sinh} \left[ x_j(l_i + N - i + \frac{1}{2}) \right] \right|}{\left| \text{sinh} \left[ x_j(N - i + \frac{1}{2}) \right] \right|} \quad;\quad l_i \equiv \sum_{m=i}^{N-1}a_m + \frac{a_N}{2}; \quad l_N \equiv \frac{a_N}{2}~,
\end{equation}
where $\left|** \right|$ is the determinant of the matrix whose rows and columns are indexed by $i$ and $j$ with $i,j = 1, 2, \ldots, N$. The $\mathcal{S}$ matrix elements\footnote{To get the elements of $\mathcal{S}$ and $\mathcal{T}$ matrices for $\mathfrak{so}(3)_k$ algebra, we need to replace $k \rightarrow \frac{k}{2}$ in eqn.(\ref{S-matrix-BN}) and eqn.(\ref{T-matrix-BN}) as discussed in \cite{Mlawer:1990uv}.} for representations $a=[a_1, \ldots, a_{N}]$ and $b=[b_1, \ldots, b_{N}]$ are given as:
\begin{equation}
\boxed{\mathcal{S}_{ab} = (-1)^{\frac{N(N-1)}{2}} \frac{2^{N-1}}{(k+2N-1)^{\frac{N}{2}}} \left| \text{sin}\left(2 \pi \frac{f_i(a) f_j(b)}{k+2N-1} \right) \right|} ~,
\label{S-matrix-BN}
\end{equation} 
where $f_i$ for an integrable representation $a$ is defined as:
\begin{equation}
f_i(a) \equiv l_i - i + \frac{2N+1}{2} ~.
\end{equation}
The quadratic matrix form elements for $\mathfrak{so}(2N+1)$ are given as:
\begin{equation}
F_{ij} =  
\begin{dcases}
    j, & j \leq i \\
    i, & i < j \leq N-1
\end{dcases}, \quad i=1,\ldots,N-1 \quad;\quad F_{iN} = \frac{i}{2}\, , \, F_{Nj} = \frac{j}{2} \, , \, F_{NN} = \frac{N}{4} ~.
\end{equation}
The $\mathcal{T}$ matrix elements can be given as:
\begin{equation}
\boxed{\mathcal{T}_{ab} = \text{exp}\left(-2 \pi i \, \frac{kN(2N+1)}{24(k+2N-1)} \right) \text{exp}\left[ \frac{2 \pi i}{4(k+2N-1)} \left( 2x_a + \frac{N}{2}y_a \right) \right]\delta_{ab}} ~,
\label{T-matrix-BN}
\end{equation}
where the integers $x_a$ and $y_a$ corresponding to a representation $a=(a_1,\ldots,a_N)$ are:
\begin{align}
x_a = \sum_{i=1}^{N-1} a_i \left( \sum_{j=1}^i ja_j + \sum_{j=i+1}^{N-1} ia_j + (2Ni-i^2 + ia_N) \right) \quad;\quad y_a = a_N(a_N + 2N) ~.
\end{align}
For the representation $a=(1,0,\ldots,0,1)$, we see that $x_a = 2N+1$. Similarly for $a=(0,\ldots,0,1)$, we have $y_a=2N+1$. Since both these representations occur for all $k \geq 2$, the integers $x_a$ and $y_a$ can be odd for certain representations. For $k=1$, the matrix can be simplified for $N \geq 3$ to:
\begin{equation}
\mathcal{T} = \text{exp}\left(-\frac{\pi i(2N+1)}{24} \right) \text{diag} \left(1, -1, \text{exp}\left[\frac{2\pi i(2N+1)}{16} \right] \right) ~.
\end{equation}
Thus, the smallest positive integer $n_0$ at which all the elements of $\mathcal{T}^{n_0}$ matrix for affine $\mathfrak{so}(2N+1)$ at level $k$ become equal can be given as:
\begin{align}
\boxed{\mathfrak{so}(3)_k}: \quad n_0 &= \begin{cases} 4, & \text{for $k=1$} \\
   4(k+2),  & \text{for $k>1$} \end{cases} \nonumber \\
		\boxed{\mathfrak{so}(2N+1)_k}: \quad n_0 &= \begin{cases}
   16, & \text{for $k=1$} \\
   8(k+2N-1),  & \text{for $k>1$, odd $N$} \\
	 4(k+2N-1),  & \text{for $k>1$, $N=2 \text{ (mod 4)}$} \\
	 2(k+2N-1),  & \text{for $k>1$, $N=0 \text{ (mod 4)}$}
	\end{cases}
  ~.
	\label{n0-for-BN}
\end{align}

\subsection*{\small{For classical Lie algebra $\boldsymbol{C_{N}=\mathfrak{sp}(2N)}$}}
The character of a representation $a$ of $\mathfrak{sp}(2N)$ where $a$ is labeled by its highest weight $a=[a_1, \ldots, a_{N}]$ and the character being evaluated at variables $e^{x_1}, e^{x_2}, \ldots, e^{x_{N}}$ is given as,
\begin{equation}
\chi_{a}[x] = \frac{\left|\text{sinh} \left[ x_j(l_i + N - i + 1) \right] \right|}{\left| \text{sinh} \left[ x_j(N - i + 1) \right] \right|} \quad;\quad l_i \equiv \sum_{m=i}^{N}a_m ~,
\end{equation}
where $\left|** \right|$ is the determinant of the matrix whose rows and columns are indexed by $i$ and $j$ with $i,j = 1, 2, \ldots, N$. Using this definition of character, the explicit expression of $\mathcal{S}$ matrix element for representations $a=[a_1, \ldots, a_{N-1}]$ and $b=[b_1, \ldots, b_{N-1}]$ can be written as: 
\begin{equation}
\boxed{\mathcal{S}_{ab} = (-1)^{\frac{N(N-1)}{2}} \left( \frac{2}{k+N+1} \right)^{\frac{N}{2}} \left| \text{sin}\left( \pi \frac{f_i(a) f_j(b)}{k+N+1} \right) \right|} ~,
\end{equation} 
where $f_i$ for an integrable representation $a$ is defined as:
\begin{equation}
f_i(a) \equiv l_i - i + N + 1 ~.
\end{equation}
The quadratic form elements for $\mathfrak{sp}(2N)$ are given as:
\begin{equation}
F_{ij} =  
\begin{dcases}
    j/2, & j=1,2,\dots,i \\
    i/2, & j=i+1,\dots,N
\end{dcases}, \quad i=1,2,\ldots,N ~.
\end{equation}
The $\mathcal{T}$ matrix elements are:
\begin{equation}
\boxed{\mathcal{T}_{ab} = \text{exp}\left(-2 \pi i \, \frac{kN(2N+1)}{24(k+N+1)} \right) \text{exp}\left( \frac{2 \pi i}{2(k+N+1)} \frac{x_a}{2} \right)\delta_{ab}} ~,
\end{equation}
where the integer $x_a$ can be given in terms of Dynkin labels of representation $a$ as:
\begin{align}
x_a = \sum_{i=1}^{N} a_i \left( \sum_{j=1}^i ja_j + \sum_{j=i+1}^{N} ia_j + i(2N+1-i) \right) ~.
\end{align}
From here, we checked for various possible $k$ and $N$ values and find that the smallest positive integer $n_0$ required such that all the elements of $\mathcal{T}^{n_0}$ matrix become equal can be given as: 
\begin{equation}
\boxed{\mathfrak{sp}(2N)_{k \geq 1}}: \quad n_0 = \begin{cases} 4, & \text{for $\mathfrak{sp}(2)$ and $k=1$} \\
   4(k+N+1),  & \text{otherwise} \end{cases} ~.
\label{n0-for-CN}
\end{equation}

\subsection*{\small{For classical Lie algebra $\boldsymbol{D_{N}=\mathfrak{so}(2N)}$}}
The character of a representation $a=[a_1, \ldots, a_{N}]$ of $\mathfrak{so}(2N)$ evaluated at variables $e^{x_1}, e^{x_2}, \ldots, e^{x_{N}}$ is given as,
\begin{align}
\chi_{a}[x] &= \frac{\left| \text{cosh} \left[ x_j(l_i + N - i) \right] \right| + \left| \text{sinh} \left[ x_j(l_i + N - i) \right] \right|}{\left| \text{cosh} \left[ x_j(N - i ) \right] \right|}; \nonumber \\
l_i &\equiv \sum_{m=i}^{N-2}a_m + \frac{a_N+a_{N-1}}{2}; \quad l_{N-1} \equiv \frac{a_N+a_{N-1}}{2}; \quad l_N \equiv \frac{a_{N}-a_{N-1}}{2}~,
\end{align}
where $\left|** \right|$ is the determinant of the matrix whose rows and columns are indexed by $i$ and $j$ with $i,j = 1, 2, \ldots, N$. Using this the expression of $\mathcal{S}$ matrix element for representations $a=[a_1, \ldots, a_{N}]$ and $b=[b_1, \ldots, b_{N}]$ can be evaluated as:
\begin{equation}
\boxed{\mathcal{S}_{ab} = (-1)^{\frac{N(N-1)}{2}} \frac{2^{N-2}}{(k+2N-2)^{\frac{N}{2}}} \left(\, \left| \text{cos}\left(\frac{2 \pi f_i(a) f_j(b)}{k+2N-2} \right) \right| + i^N \left| \text{sin}\left(\frac{2 \pi f_i(a) f_j(b)}{k+2N-2} \right) \right| \,\right)} ~,
\end{equation} 
where $f_i$ for the integrable representation $a$ is given as:
\begin{equation}
f_i(a) \equiv l_i - i + N ~.
\end{equation}
The quadratic form for $\mathfrak{so}(2N)$ are given as:
\begin{align}
& F_{ij}  =  
\begin{dcases}
    j, & j=1,2,\ldots,i \\
    i, & j=i+1,\ldots,N-2
\end{dcases}\quad,\quad i=1,2,\ldots,N-2 \nonumber \\
& F_{N-1,j}  = F_{Nj} = \frac{j}{2} \quad;\quad F_{i,N-1} = F_{iN} = \frac{i}{2} \quad \, (i,j=1,\ldots,N-2) \nonumber \\
& F_{N-1,N-1} = F_{NN} = \frac{N}{4} \quad;\quad F_{N,N-1} = F_{N-1,N} = \frac{N-2}{4} ~.
\end{align}
Using quadratic form, the $\mathcal{T}$ matrix elements are:
\begin{equation}
\boxed{\mathcal{T}_{ab} = \text{exp}\left(-2 \pi i \, \frac{kN(2N-1)}{24(k+2N-2)} \right) \text{exp}\left[ \frac{2 \pi i}{2(k+2N-2)} \left( x_a + \frac{N}{4}y_a - a_N a_{N-1} \right) \right]\delta_{ab}} ~,
\end{equation}
where the integers $x_a$ and $y_a$ for a representation $a=(a_1,\ldots,a_N)$ are given as:
\begin{align}
x_a &= \sum_{i=1}^{N-2} a_i \left( \sum_{j=1}^i ja_j + \sum_{j=i+1}^{N-2} ia_j + (2Ni-3i-i^2) \right) + (a_{N-1}+a_N+2)\sum_{j=1}^{N-2} ja_j \nonumber \\
y_a &= (a_N+a_{N-1})^2 + 2(N-1)(a_N + a_{N-1}) ~.
\end{align}
For representation $a=(1,0,\ldots,0)$, we get $x_a = 2N-3$ and for rep $a=(0,\ldots,0,1)$, one obtains $y_a = 2N-1$. Since these two representations are always present for any $k \geq 1$, we can say that the integers $x_a$ and $y_a$ can be odd integers. For $k=1$, the matrix elements can be simplified for $N \geq 4$ as:
\begin{equation}
\mathcal{T}_{ab} = \text{exp}\left(-\frac{\pi i N}{12} \right) \text{diag}\left(1, -1, \text{exp}\left(\frac{\pi i N}{4}\right), \text{exp}\left(\frac{\pi i N}{4}\right) \right) ~.
\end{equation}
Thus, the smallest positive integer $n_0$ at which the $\mathcal{T}^{n_0}$ matrix for affine $\mathfrak{so}(2N)$ at level $k$ becomes a scalar matrix can be given as:
\begin{align}
\boxed{\mathfrak{so}(2N)_{k=1}}&: \quad n_0 = \begin{cases}
   8,  & \text{for $N=$ odd} \\
	 4,  & \text{for $N=2 \text{ (mod 4)}$} \\
	 2,  & \text{for $N=0 \text{ (mod 4)}$}
	\end{cases} \nonumber \\
\boxed{\mathfrak{so}(2N)_{k>1}}&: \quad n_0 = {\begin{cases}
   8(k+2N-2),  & \text{for odd $N$} \\
	 4(k+2N-2),  & \text{for $N=2 \text{ (mod 4)}$} \\
	 2(k+2N-2),  & \text{for $N=0 \text{ (mod 4)}$}
	\end{cases}}  
  ~.
	\label{n0-for-DN}
\end{align}
}

\vspace{0.5cm}
\textbf{Acknowledgements} 
We would like to thank R.K. Kaul, A. Mironov, Andrey Morozov, Satoshi Nawata and A. Sleptsov for discussion and correspondence. Vivek would like to thank Y. Kononov, Vinay Malvimat, Sudip Ghosh, Pankaj Mandal, Pranjal Nayak, Ronak and Priyanka for helpful discussion. Ramadevi would like to thank MPIM Bonn for local hospitality where initial stages of the work were done. Ramadevi, Vivek and Saswati would like to thank DST-RFBR grant(INT/RUS/RFBR/P-231) for the support. Saswati would like to thank CSIR for the research fellowship.
\bibliographystyle{JHEP}
\bibliography{EE-for-Links} 

\providecommand{\href}[2]{#2}\begingroup\raggedright\begin{thebibliography}{10}

\bibitem{Ryu:2006bv}
S.~Ryu and T.~Takayanagi, \emph{{Holographic derivation of entanglement entropy
  from AdS/CFT}},
  \href{https://doi.org/10.1103/PhysRevLett.96.181602}{\emph{Phys. Rev. Lett.}
  {\bfseries 96} (2006) 181602},
  [\href{https://arxiv.org/abs/hep-th/0603001}{{\ttfamily hep-th/0603001}}].

\bibitem{Hubeny:2007xt}
V.~E. Hubeny, M.~Rangamani and T.~Takayanagi, \emph{{A Covariant holographic
  entanglement entropy proposal}},
  \href{https://doi.org/10.1088/1126-6708/2007/07/062}{\emph{JHEP} {\bfseries
  07} (2007) 062}, [\href{https://arxiv.org/abs/0705.0016}{{\ttfamily
  0705.0016}}].

\bibitem{Lewkowycz:2013nqa}
A.~Lewkowycz and J.~Maldacena, \emph{{Generalized gravitational entropy}},
  \href{https://doi.org/10.1007/JHEP08(2013)090}{\emph{JHEP} {\bfseries 08}
  (2013) 090}, [\href{https://arxiv.org/abs/1304.4926}{{\ttfamily 1304.4926}}].

\bibitem{Bianchi:2012ev}
E.~Bianchi and R.~C. Myers, \emph{{On the Architecture of Spacetime Geometry}},
  \href{https://doi.org/10.1088/0264-9381/31/21/214002}{\emph{Class. Quant.
  Grav.} {\bfseries 31} (2014) 214002},
  [\href{https://arxiv.org/abs/1212.5183}{{\ttfamily 1212.5183}}].

\bibitem{Balasubramanian:2013rqa}
V.~Balasubramanian, B.~Czech, B.~D. Chowdhury and J.~de~Boer, \emph{{The
  entropy of a hole in spacetime}},
  \href{https://doi.org/10.1007/JHEP10(2013)220}{\emph{JHEP} {\bfseries 10}
  (2013) 220}, [\href{https://arxiv.org/abs/1305.0856}{{\ttfamily 1305.0856}}].

\bibitem{Balasubramanian:2013lsa}
V.~Balasubramanian, B.~D. Chowdhury, B.~Czech, J.~de~Boer and M.~P. Heller,
  \emph{{Bulk curves from boundary data in holography}},
  \href{https://doi.org/10.1103/PhysRevD.89.086004}{\emph{Phys. Rev.}
  {\bfseries D89} (2014) 086004},
  [\href{https://arxiv.org/abs/1310.4204}{{\ttfamily 1310.4204}}].

\bibitem{Myers:2014jia}
R.~C. Myers, J.~Rao and S.~Sugishita, \emph{{Holographic Holes in Higher
  Dimensions}}, \href{https://doi.org/10.1007/JHEP06(2014)044}{\emph{JHEP}
  {\bfseries 06} (2014) 044},
  [\href{https://arxiv.org/abs/1403.3416}{{\ttfamily 1403.3416}}].

\bibitem{Swingle:2009bg}
B.~Swingle, \emph{{Entanglement Renormalization and Holography}},
  \href{https://doi.org/10.1103/PhysRevD.86.065007}{\emph{Phys. Rev.}
  {\bfseries D86} (2012) 065007},
  [\href{https://arxiv.org/abs/0905.1317}{{\ttfamily 0905.1317}}].

\bibitem{Nozaki:2013vta}
M.~Nozaki, T.~Numasawa, A.~Prudenziati and T.~Takayanagi, \emph{{Dynamics of
  Entanglement Entropy from Einstein Equation}},
  \href{https://doi.org/10.1103/PhysRevD.88.026012}{\emph{Phys. Rev.}
  {\bfseries D88} (2013) 026012},
  [\href{https://arxiv.org/abs/1304.7100}{{\ttfamily 1304.7100}}].

\bibitem{Lashkari:2013koa}
N.~Lashkari, M.~B. McDermott and M.~Van~Raamsdonk, \emph{{Gravitational
  dynamics from entanglement 'thermodynamics'}},
  \href{https://doi.org/10.1007/JHEP04(2014)195}{\emph{JHEP} {\bfseries 04}
  (2014) 195}, [\href{https://arxiv.org/abs/1308.3716}{{\ttfamily 1308.3716}}].

\bibitem{Faulkner:2013ica}
T.~Faulkner, M.~Guica, T.~Hartman, R.~C. Myers and M.~Van~Raamsdonk,
  \emph{{Gravitation from Entanglement in Holographic CFTs}},
  \href{https://doi.org/10.1007/JHEP03(2014)051}{\emph{JHEP} {\bfseries 03}
  (2014) 051}, [\href{https://arxiv.org/abs/1312.7856}{{\ttfamily 1312.7856}}].

\bibitem{Swingle:2014uza}
B.~Swingle and M.~Van~Raamsdonk, \emph{{Universality of Gravity from
  Entanglement}},  \href{https://arxiv.org/abs/1405.2933}{{\ttfamily
  1405.2933}}.

\bibitem{Calabrese:2004eu}
P.~Calabrese and J.~L. Cardy, \emph{{Entanglement entropy and quantum field
  theory}}, \href{https://doi.org/10.1088/1742-5468/2004/06/P06002}{\emph{J.
  Stat. Mech.} {\bfseries 0406} (2004) P06002},
  [\href{https://arxiv.org/abs/hep-th/0405152}{{\ttfamily hep-th/0405152}}].

\bibitem{Calabrese:2009qy}
P.~Calabrese and J.~Cardy, \emph{{Entanglement entropy and conformal field
  theory}}, \href{https://doi.org/10.1088/1751-8113/42/50/504005}{\emph{J.
  Phys.} {\bfseries A42} (2009) 504005},
  [\href{https://arxiv.org/abs/0905.4013}{{\ttfamily 0905.4013}}].

\bibitem{Klebanov:2011uf}
I.~R. Klebanov, S.~S. Pufu, S.~Sachdev and B.~R. Safdi, \emph{{Renyi Entropies
  for Free Field Theories}},
  \href{https://doi.org/10.1007/JHEP04(2012)074}{\emph{JHEP} {\bfseries 04}
  (2012) 074}, [\href{https://arxiv.org/abs/1111.6290}{{\ttfamily 1111.6290}}].

\bibitem{Casini:2009sr}
H.~Casini and M.~Huerta, \emph{{Entanglement entropy in free quantum field
  theory}}, \href{https://doi.org/10.1088/1751-8113/42/50/504007}{\emph{J.
  Phys.} {\bfseries A42} (2009) 504007},
  [\href{https://arxiv.org/abs/0905.2562}{{\ttfamily 0905.2562}}].

\bibitem{Fursaev:2012mp}
D.~V. Fursaev, \emph{{Entanglement Renyi Entropies in Conformal Field Theories
  and Holography}}, \href{https://doi.org/10.1007/JHEP05(2012)080}{\emph{JHEP}
  {\bfseries 05} (2012) 080},
  [\href{https://arxiv.org/abs/1201.1702}{{\ttfamily 1201.1702}}].

\bibitem{Witten:1988hf}
E.~Witten, \emph{{Quantum Field Theory and the Jones Polynomial}},
  \href{https://doi.org/10.1007/BF01217730}{\emph{Commun. Math. Phys.}
  {\bfseries 121} (1989) 351--399}.

\bibitem{Levin:2006zz}
M.~Levin and X.-G. Wen, \emph{{Detecting Topological Order in a Ground State
  Wave Function}},
  \href{https://doi.org/10.1103/PhysRevLett.96.110405}{\emph{Phys. Rev. Lett.}
  {\bfseries 96} (2006) 110405}.

\bibitem{Dong:2008ft}
S.~Dong, E.~Fradkin, R.~G. Leigh and S.~Nowling, \emph{{Topological
  Entanglement Entropy in Chern-Simons Theories and Quantum Hall Fluids}},
  \href{https://doi.org/10.1088/1126-6708/2008/05/016}{\emph{JHEP} {\bfseries
  05} (2008) 016}, [\href{https://arxiv.org/abs/0802.3231}{{\ttfamily
  0802.3231}}].

\bibitem{Kitaev:2005dm}
A.~Kitaev and J.~Preskill, \emph{{Topological entanglement entropy}},
  \href{https://doi.org/10.1103/PhysRevLett.96.110404}{\emph{Phys. Rev. Lett.}
  {\bfseries 96} (2006) 110404},
  [\href{https://arxiv.org/abs/hep-th/0510092}{{\ttfamily hep-th/0510092}}].

\bibitem{Balasubramanian:2014hda}
V.~Balasubramanian, P.~Hayden, A.~Maloney, D.~Marolf and S.~F. Ross,
  \emph{{Multiboundary Wormholes and Holographic Entanglement}},
  \href{https://doi.org/10.1088/0264-9381/31/18/185015}{\emph{Class. Quant.
  Grav.} {\bfseries 31} (2014) 185015},
  [\href{https://arxiv.org/abs/1406.2663}{{\ttfamily 1406.2663}}].

\bibitem{Marolf:2015vma}
D.~Marolf, H.~Maxfield, A.~Peach and S.~F. Ross, \emph{{Hot multiboundary
  wormholes from bipartite entanglement}},
  \href{https://doi.org/10.1088/0264-9381/32/21/215006}{\emph{Class. Quant.
  Grav.} {\bfseries 32} (2015) 215006},
  [\href{https://arxiv.org/abs/1506.04128}{{\ttfamily 1506.04128}}].

\bibitem{Salton:2016qpp}
G.~Salton, B.~Swingle and M.~Walter, \emph{{Entanglement from Topology in
  Chern-Simons Theory}},
  \href{https://doi.org/10.1103/PhysRevD.95.105007}{\emph{Phys. Rev.}
  {\bfseries D95} (2017) 105007},
  [\href{https://arxiv.org/abs/1611.01516}{{\ttfamily 1611.01516}}].

\bibitem{Balasubramanian:2016sro}
V.~Balasubramanian, J.~R. Fliss, R.~G. Leigh and O.~Parrikar,
  \emph{{Multi-Boundary Entanglement in Chern-Simons Theory and Link
  Invariants}}, \href{https://doi.org/10.1007/JHEP04(2017)061}{\emph{JHEP}
  {\bfseries 04} (2017) 061},
  [\href{https://arxiv.org/abs/1611.05460}{{\ttfamily 1611.05460}}].

\bibitem{Aganagic:2011sg}
M.~Aganagic and S.~Shakirov, \emph{{Knot Homology and Refined Chern-Simons
  Index}}, \href{https://doi.org/10.1007/s00220-014-2197-4}{\emph{Commun. Math.
  Phys.} {\bfseries 333} (2015) 187--228},
  [\href{https://arxiv.org/abs/1105.5117}{{\ttfamily 1105.5117}}].

\bibitem{fulton2013representation}
W.~Fulton and J.~Harris, \emph{Representation theory: a first course},
  vol.~129.
\newblock Springer Science \& Business Media, 2013.

\bibitem{eliahou2003infinite}
S.~Eliahou, L.~H. Kauffman and M.~B. Thistlethwaite, \emph{Infinite families of
  links with trivial jones polynomial}, {\emph{Topology} {\bfseries 42} (2003)
  155--169}.

\bibitem{morton1996distinguishing}
H.~R. Morton and P.~R. Cromwell, \emph{Distinguishing mutants by knot
  polynomials}, {\emph{Journal of Knot Theory and its Ramifications} {\bfseries
  5} (1996) 225--238}.

\bibitem{Nawata:2015xha}
S.~Nawata, P.~Ramadevi and V.~K. Singh, \emph{{Colored HOMFLY polynomials that
  distinguish mutant knots}},
  \href{https://arxiv.org/abs/1504.00364}{{\ttfamily 1504.00364}}.

\bibitem{Mironov:2015aia}
A.~Mironov, A.~Morozov, A.~Morozov, P.~Ramadevi and V.~K. Singh, \emph{{Colored
  HOMFLY polynomials of knots presented as double fat diagrams}},
  \href{https://doi.org/10.1007/JHEP07(2015)109}{\emph{JHEP} {\bfseries 07}
  (2015) 109}, [\href{https://arxiv.org/abs/1504.00371}{{\ttfamily
  1504.00371}}].

\bibitem{Mlawer:1990uv}
E.~J. Mlawer, S.~G. Naculich, H.~A. Riggs and H.~J. Schnitzer, \emph{{Group
  level duality of WZW fusion coefficients and Chern-Simons link observables}},
  \href{https://doi.org/10.1016/0550-3213(91)90110-J}{\emph{Nucl. Phys.}
  {\bfseries B352} (1991) 863--896}.

\bibitem{francesco2012conformal}
P.~Francesco, P.~Mathieu and D.~S{\'e}n{\'e}chal, \emph{Conformal field
  theory}.
\newblock Springer Science \& Business Media, 2012.

\end{thebibliography}\endgroup

\end{document}